\documentclass[journal]{IEEEtran}
\pdfoutput=1
\usepackage[pdftex]{graphicx}
\DeclareGraphicsExtensions{.pdf,.jpeg,.png}
\usepackage[cmex10]{amsmath}
\usepackage{amsfonts, amssymb}
\usepackage{amssymb} 
\usepackage{cite} 
\usepackage[ruled,norelsize]{algorithm2e} 
\usepackage{multirow}
\makeatletter
\newcommand{\removelatexerror}{\let\@latex@error\@gobble}
\makeatother
\makeatletter

\begin{document} 
\newcommand{\wt}[2]{W_{#1}[#2]}
\newcommand{\wh}[2]{\hat{W}_{#1}[#2]}
\newcommand{\ws}[3]{\textbf{W}_{#1;#2}^{#3}} 
\newcommand{\wsh}[3]{\hat{\textbf{W}}_{#1;#2}^{#3}}
\newcommand{\Nob}[3]{N_{ob,A_{#1}}^{(#2,#3)}}
\newcommand{\NobE}[3]{\overline{N}_{ob,A_{#1}}^{(#2,#3)}}
\newcommand{\NobTwo}[2]{N_{ob,A_{#1}}^{(#2)}}
\newcommand{\NobETwo}[2]{\overline{N}_{ob,A_{#1}}^{(#2)}}
\newcommand{\Pob}[3]{P_{ob,A_{#1}}^{(#2,#3)}}
\newcommand{\pr}{\text{Pr}} 
\newcommand{\xstar}{n}
\newcommand{\ystar}{q} 
\newcommand{\tx}{TX} 
\newcommand{\rx}{RX} 
\newcommand{\xiN}[1]{\xi_{#1}} 
\newcommand{\pe}[1]{P_{e_{#1}}} 
\newcommand{\peE}[1]{\overline{P}_{e_{#1}}} 
\newcommand{\Rk}[1]{R_{#1}} 
\newcommand{\vf}{\hspace{-0.8 mm}} 
\newcommand{\rom}[1]{\expandafter\@slowromancap\romannumeral #1@}
\title{Analysis and Design of Multi-Hop Diffusion-Based Molecular Communication Networks}
\author{Arman~Ahmadzadeh, ~\IEEEmembership{Student Member,~IEEE,}
        Adam~Noel,~\IEEEmembership{Student Member,~IEEE,}
        and~Robert~Schober,~\IEEEmembership{Fellow,~IEEE}
\thanks{This work will be presented in part at IEEE GLOBECOM 2014 \cite{Arman1}.}        
\thanks{A. Ahmadzadeh and R. Schober are with the Institute for Digital Communication, University of Erlangen-Nuremberg, Erlangen, Germany 
		(email: \{ahmadzadeh, schober\}@LNT.de)}
\thanks{A. Noel is with the Department of Electrical and Computer Engineering, University of British Columbia, Vancuver, BC, Canada, V6T 1Z4 (email: adamn@ece.ubc.ca) }}
\maketitle

\begin{abstract}
In this paper, we consider a multi-hop molecular communication network consisting of one nanotransmitter, one nanoreceiver, and multiple nanotransceivers acting as relays. We consider three different relaying schemes to improve the range of diffusion-based molecular communication. In the first scheme, different types of messenger molecules are utilized in each hop of the multi-hop network. In the second and third scheme, we assume that two types of molecules and one type of molecule are utilized in the network, respectively. We identify self-interference, backward intersymbol interference (backward-ISI), and forward-ISI as the performance-limiting effects for the second and third relaying schemes. Furthermore, we consider two relaying modes analogous to those used in wireless communication systems, namely full-duplex and half-duplex relaying. We propose the adaptation of the decision threshold as an effective mechanism to mitigate self-interference and backward-ISI at the relay for full-duplex and half-duplex transmission. We derive closed-form expressions for the expected end-to-end error probability of the network for the three considered relaying schemes. Furthermore, we derive closed-form expressions for the optimal number of molecules released by the nanotransmitter and the optimal detection threshold of the nanoreceiver for minimization of the expected error probability of each hop.              
\end{abstract}

\IEEEpeerreviewmaketitle
\section{Introduction}
\IEEEPARstart{R}{ecent}
advancements in the field of nanotechnology have enabled the development of small-scale devices, so-called nanomachines. Nanomachines have functional components that are on the order of nanometers in size ($10^{-9}$ m) and are only capable of performing simple computation, sensing, or actuation tasks \cite{Akyildiz1}. Their limited processing capabilities prevent single nanomachines from executing more complex tasks. Hence, it is envisioned that networks of nanomachines, so-called nanonetworks, have to be formed to perform more elaborate and challenging tasks in a distributed manner. For this to be possible, nanomachines have to be able to communicate with each other. One of the most important areas for application of nanonetworks is the biomedical domain, which includes health monitoring, tissue engineering, and targeted drug delivery. Other application domains of nanonetworks include industrial applications, such as new materials and quality control of products, and environmental applications, such as biodegradation and air pollution control; see \cite{Nakano1}, \cite{Chahibi}. 

Different approaches have been proposed for communication among nanomachines in the literature such as communication based on hard junctions, electromagnetic waves, acoustic waves, and molecular communication (MC); see \cite{Akyildiz1}. Among these different approaches, MC has the advantages of energy efficiency and potential biocompatibility. In fact, MC is already used by nature for communication among biological entities and systems, such as molecules, cells, organelles, and organisms. In MC, molecules are the carriers of information, which is in contrast to conventional wireless communication systems, where electromagnetic waves are employed for this purpose; see \cite{NakanoB}. In diffusion-based MC, the signal molecules that are released by the transmitter nanomachine in a fluid environment randomly ``walk'' in all directions without any further infrastructure and some of them may reach the receiver nanomachine. Diffusion-based MC has been extensively  studied in the literature, cf. e.g. \cite{Pierobon1, ShahMohammadian1, Kadloor1, Kuran1}. MC nanonetworks for nanomachines pose unique challenges that are not commonly found in traditional communication networks and these challenges have to be taken into account in the development of practical communication protocols for such networks. One of the challenges in MC is that the propagation time increases with the square of the distance. If an intended receiver is far away from the transmitter, then using a single transmitter may be impractical. One approach in conventional wireless communications that can be adapted for MC is the use of intermediate transceivers acting as relays to aid the communication with distant receivers. Such relays can potentially improve the reliability and performance of a communication link.

In fact, relaying of information also plays an important role in communication among biological systems. For example, in typical communication between cells, a \textit{signaling cell} produces a particular type of signal molecule that is detected by a \textit{target cell}; see \cite{AlbertsBook}. The target cell possesses \textit{receptor proteins} that recognize and respond specifically to the signal molecule. If a signal molecule is detected by a cell-surface receptor, then this information is relayed into the interior of the target cell via a set of \textit{intra-cellular signaling molecules}, which act in sequence and ultimately change the behaviour of the target cell. The reception and transduction of signal molecules is called cell signaling. The components of this intracellular relay system perform crucial functions that are similar to the processing in conventional wireless relaying schemes, such as decode-and-forward (DF) and amplify-and-forward (AF) \cite{hossain2011cooperative}. In the following, we provide two examples of such components \cite[Ch. 16]{AlbertsBook}: 
\begin{enumerate}
	\item Cell receptor proteins act like a DF relay. In the presence of extra-cellular signal molecules, specific signal molecules bind to receptors, i.e., they decode the message. Then, they forward the message by either opening an ion channel, e.g., ion-channel-coupled receptors, in the plasma membrane, or activate another protein or enzyme in the intra-cellular signaling pathway, e.g., G-protein-coupled receptors (GPCRs) or enzyme-coupled receptors. 
	\item Once the GPCRs are activated by binding to the first messenger molecule, i.e., an extra-cellular molecule, the G-protein, in turn, activates target enzymes such as \textit{adenylyl cyclase} or \textit{phospholipase C}. These two target enzymes act like AF relays. When they are activated, they produce large numbers of small intra-cellular molecules, in other words, they amplify the message. Increasing the amount of small intra-cellular molecules such as \textit{cyclic AMP} or \textit{phospholipase} regulates the activation of other proteins in the intra-cellular signaling pathway and leads to changes in the behaviour of the target cell.           
\end{enumerate} 

Several works have recently addressed multi-hop communication among nanomachines. Network layer issues in multi-hop nanonetworks have been introduced in \cite{Akyildiz1}, \cite{Nakano1}. In \cite{Einolghozati1}, \cite{Einolghozati2}, a diffusion-based multi-hop network among bacteria colonies was analyzed, where each node of the network was formed by a population of bacteria. Two relaying schemes, namely sense-and-forward and DF, were proposed, and the resulting improvements in channel capacity compared to the no-relay case were confirmed by simulations. In \cite{Nakano2}, \cite{Nakano3}, the design and analysis of repeater cells in Calcium junction channels, where signal molecules propagate through repetitive processes of diffusion and amplification, were investigated. In \cite{Unluturk1}, the rate-delay trade-off of a three-node nanonetwork for a specific messenger molecule, polyethylene, was analyzed for network coding at the relay node. A multi-hop routing mechanism, where bacteria are deployed as information carriers, was analyzed in \cite{Balasubramaniam1}. In \cite{Balasubramaniam1}, the authors adapted some of the features of bacteria such as conjugation and chemotaxis to mimic mechanisms found in internet protocol (IP)-based communication networks, such as opportunistic routing, packet addressing, and packet filtering. The authors in \cite{Walsh1} proposed the use of virus particles as information carriers, where the information was encoded in the Deoxyribonucleic Acid (\textit{DNA}) or Ribonucleic Acid (\textit{RNA}) part of the virus particle. Numerical results showed that the reliability of the proposed network improved by increasing the number of intermediate nodes at the cost of increasing the total delay of transmission. Despite these prior works, to the best of the authors' knowledge, the effects of \emph{multiple} transmissions of \emph{random} symbols (or bits) on the performance of MC based multi-hop networks and corresponding mitigation techniques have not been studied in the literature so far.

In this paper, we assume that the transmitter nanomachine emits multiple random bits, and we investigate three different relaying schemes, namely multi-molecule multi-hop (MM-MH), two-molecule multi-hop (2M-MH), and single-molecule multi-hop (SM-MH). In MM-MH, we assume that a different type of molecule is used in each hop of the network. In 2M-MH and SM-MH, we assume that the number of available molecule types to be used as information carriers is limited. In particular, in 2M-MH, we assume that two different types of molecules are available, and in SM-MH, only one type of molecule is available for use at all relays of the network for transmission and detection. Hence, in 2M-MH and SM-MH, multiple transmissions of random bits by the transmitter nanomachine lead to the occurrence of self-interference, backward-ISI, and forward-ISI. Self-interference occurs when one relay must detect the same type of molecule that it also emits. This effect has also been mentioned in \cite{Einolghozati3}, however, has not been considered in the analysis. Backward-ISI and forward-ISI occur more generally when the same type of molecule is used for transmission by multiple nodes of the network. On the other hand, in MM-MH interference is avoided completely. However, in practice, it might be difficult to find a sufficient number of unique molecule types to use in each hop, such that only the intended receiver is sensitive to the corresponding type of molecule. Furthermore, with many different types of molecules in use, it becomes more likely that some of the used types of molecules cause interference to adjacent communication links employing the same types of molecules as the carrier of information or cause inadvertent reactions elsewhere in the environment. Hence, 2M-MH and SM-MH are better suited for use in environments with restrictions on the types of deployed molecules. However, their performance is limited by interference and effective mitigation techniques are needed.     

This paper expands the work presented in \cite{Arman1} and makes the following contributions: 
\begin{enumerate}
	\item As in \cite{Arman1}, we derive a closed-form analytical expressions for the expected error probability of two-hop networks. We expand these expressions to the multi-hop scenario for the three above-mentioned relaying schemes based on the error analysis of a single link described in \cite{NoelJ1}, \cite{NoelJ2}. 
	\item For a single link, we minimize the expected error probability by deriving closed-form analytical expressions for the optimal detection threshold at the receiver side and the optimal number of  molecules released by the transmitter.
	\item As in \cite{Arman1}, we identify the impact of self-interference in a two-hop network and propose two techniques to mitigate it. In the first method, the relay adjusts its detection threshold in each bit interval based on all previously-detected information bits at the relay. In the second method, half-duplex relaying is employed instead of full-duplex relaying. 
	\item We identify the performance-limiting effects of backward-ISI and forward-ISI for 2M-MH, and propose the adaptation of the decision threshold as a means to mitigate backward-ISI. We combine self-interference and backward-ISI mitigation to cope with the detrimental effects of both types of interference in SM-MH.      
\end{enumerate}  

The rest of this paper is organized as follows. In Section \ref{Sec.SysModandPre}, we introduce the system model and the preliminaries for the error rate analysis. In Sections \ref{Sec.Two-hopNetPerAna}, \ref{Sec. TM-MH}, and \ref{Sec. SM-MH}, we evaluate the expected error probabilities of MM-MH, 2M-MH, and SM-MH, respectively. Numerical results are given in Section \ref{Sec.NumRes}, and conclusions are drawn in Section \ref{Sec.Con}. 

\section{SYSTEM MODEL AND PRELIMINARIES} 
\label{Sec.SysModandPre} 
In this section, we introduce the system model and some preliminaries regarding the error rate analysis of a single link, which we require as a prerequisite for the error rate analysis of the multi-hop network. 
\subsection{System Model}
\begin{figure}[!t]
	\centering
	\includegraphics[scale = 0.57]{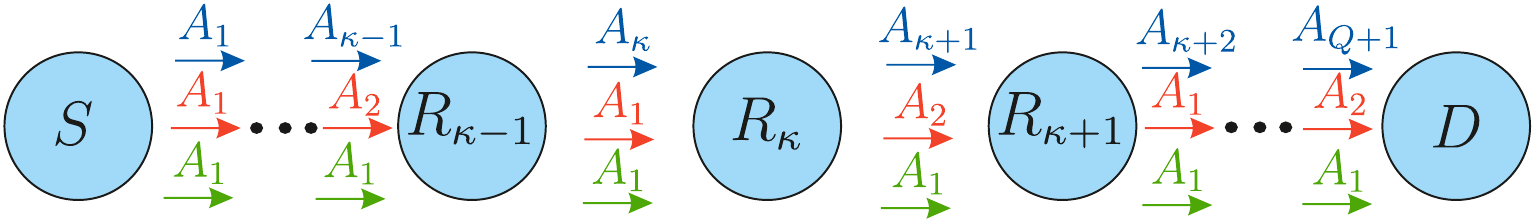}
	\caption{System model of a multi-hop MC network, where the molecules used for MM-MH, 2M-MH, and SM-MH are shown in blue, red, and green, respectively.}
	\label{Fig.SysMod}
\end{figure} 
In this paper, we use the terms ``nanomachine'' and ``node'' interchangeably to refer to the devices in the network, as the term ``node" is commonly used in the relaying literature. We assume that a source ($S$) node and a destination ($D$) node are placed at locations $(0,0,0)$ and $(x_{D},0,0)$ of a 3-dimensional space, respectively. We assume that there are $Q$ relay nodes, $Q \in \mathbb{Z}^+$, and the $\kappa$th relay ($\Rk{\kappa}$) node is placed at $(x_{\kappa},0,0)$, $\kappa \in \lbrace 1,2,...,Q \rbrace$, along the $x$-axis. The relays are equally spaced between node $S$ and node $D$, i.e., $x_{\kappa} = \kappa x_{D} / (Q + 1) $, cf. Fig. \ref{Fig.SysMod}. We assume that nodes $D$ and $\Rk{\kappa}$ are spherical in shape with fixed volumes (and radii) $V_{D}$ $(r_{D})$ and $V_{R_{\kappa}}$ $(r_{R_{\kappa}})$, respectively, and that they are passive observers such that molecules can diffuse through them without reacting. In this paper, for convenience of notation, we also refer to node $S$ and node $D$ as $\Rk{0}$ and $\Rk{Q+1}$, respectively.   

We consider three different relaying schemes, namely MM-MH, 2M-MH, and SM-MH, where the number of available types of molecules, the type of detected molecules at relay node $\Rk{\kappa}$, and the type of emitted molecules at relay node $\Rk{\kappa}$ for each relaying scheme are given in Table \ref{Table1}.

\begin{table}[t]
\renewcommand{\arraystretch}{1}
\caption{Properties of the Relaying Schemes}
\label{Table1}
\centering
\begin{tabular}{|c|c|c|c|}
\hline 
\bfseries Relaying & \bfseries \# of Available Types  & \multicolumn{2}{|c|}{\bfseries Relay Node $\Rk{\kappa}$} \\ \cline{3-4}
\bfseries Scheme & \bfseries  of Molecules & \bfseries Detects & \bfseries Emits \\ 
\hline 
MM-MH & $Q+1$ & $A_{\kappa}$ & $A_{\kappa+1}$ \\ 
\hline 
2M-MH & 2 & $A_1 (A_2)$ & $A_2 (A_1)$ \\ 
\hline 
SM-MH & 1 & $A_1$ & $A_1$ \\  
\hline 
\end{tabular}
\end{table} 

Furthermore, we assume that the information that is sent from node $S$ to node $D$ is encoded into a binary sequence of length $K$, $\textbf{W}_{S} = \{ \wt{S}{1}, \wt{S}{2},..., \wt{S}{K} \}$. Here, $\wt{S}{j}$ is the bit transmitted by node $S$ in the $j$th bit interval with $\pr(\wt{S}{j}=1)=P_{1}$, and $\pr(\wt{S}{j}=0)=P_{0}=1-P_{1}$, where $\pr(\cdot)$ denotes probability. The information bits transmitted and detected by relay $\Rk{\kappa}$ in the $j$th bit interval are denoted by $\wt{\Rk{\kappa}}{j}$ and $\wh{\Rk{\kappa}}{j}$, respectively. The information bit detected at node $D$ in the $j$th bit interval is denoted by $\wh{D}{j}$. In the following, we denote a sequence of bits transmitted and detected by node $h$, $h \in \{S,\Rk{1},...,\Rk{\kappa},D\}$ by $\ws{h}{a}{b} = \{ \wt{h}{a},...,\wt{h}{b} \}$ and $ \wsh{h}{a}{b} = \{ \wh{h}{a},...,\wh{h}{b} \}$, respectively. We adopt ON/OFF keying for modulation and a fixed bit interval duration of $T$ seconds. This is a commonly-used modulation scheme in the MC literature; cf. e.g. \cite{NoelJ2 ,Nakano4, NoelJ3, Atakan1, Llatser1}. The number of released molecules of type $A_{f}, f \in \{ 1,2,...,Q+1\}$, released by a transmitting node at the beginning of a bit interval to convey information bit ``1'' is denoted as $N_{A_{f}}$. No molecules are released to convey information bit ``0''. The concentration of type $A_{f}$ molecules at the point defined by vector $\vec{r}$ at time $t$ in molecule $\cdot$ $\text{m}^{-3}$ is denoted by $C_{A_{f}}( \vec{r}, t )$. We assume that the movements of individual molecules are independent. 

We adopt the DF-relaying protocol, where the relay first decodes the received message, and then re-encodes the detected message for re-transmission. Furthermore, we consider two relaying modes for multi-hop transmission, namely full-duplex and half-duplex. For full-duplex transmission, reception and transmission occur simultaneously at the relay node, i.e., in each bit interval, relay $\Rk{\kappa}$ detects the information transmitted by node $\Rk{\kappa - 1}$, and forwards the information bit detected in the previous bit interval to node $\Rk{\kappa + 1}$. For half-duplex transmission, the relay performs detection and reception separately, i.e., in one bit interval, relay $\Rk{\kappa}$ detects the information transmitted by node $\Rk{\kappa - 1}$, and in the next bit interval, relay $\Rk{\kappa}$ forwards the detected information bit to node $\Rk{\kappa + 1}$. In a multi-hop network consisting of $Q$ relays and using either protocol, relay $\Rk{\kappa}$ detects the first bit in the $\kappa$th bit interval. In other words, it is silent in the first $\kappa$ bit intervals and does not transmit. Thus, the total duration required for the transmission of $L$ bits of information is $KT$, where $K=L+Q$ and $K=2L+Q-1$ for full-duplex and half-duplex relaying, respectively. We note that in the sequence transmitted by node $\Rk{\kappa}$, i.e., $\ws{\Rk{\kappa}}{1}{K}$, the first $\kappa$ and last $Q-\kappa$ bits are zero. Furthermore, for half-duplex relaying, we assume that node $\Rk{\kappa}$, where $\kappa$ is even (odd), transmits in odd (even) bit intervals, i.e., $\wt{\Rk{\kappa}}{2i+\kappa}$ $(\wt{\Rk{\kappa}}{2i+\kappa+1}), i \in \{ 0,...,L-1 \}$. In even (odd) bit intervals, node $\Rk{\kappa}$ is silent, i.e., $\wt{\Rk{\kappa}}{2i+\kappa+1}=0$ $(\wt{\Rk{\kappa}}{2i+\kappa}=0), i \in \{ 1,...,L-1 \}$, and node $\Rk{\kappa+1}$ does not detect. For example, a sequence transmitted by node $\Rk{\kappa}$ for full-duplex and half duplex relaying when $\kappa$ is odd is given by (\ref{Eq. SequenceTranFD}) and (\ref{Eq. SequenceTranHD}), respectively.
\begin{figure*} 
\begin{equation}
	\label{Eq. SequenceTranFD} 
	\ws{\Rk{\kappa}}{1}{K}  =  \left\lbrace \underbrace{0,\vf\cdots\vf ,0}_\text{First $\kappa$ bits}, \underbrace{ \wt{\Rk{\kappa}}{\kappa \vf + \vf 1}, \wt{\Rk{\kappa}}{\kappa \vf + \vf 2},\cdots , \wt{\Rk{\kappa}}{\kappa \vf + \vf L}, }_\text{$L$ bits of information}
	 \underbrace{0, \cdots, 0}_\text{Last $Q-\kappa$ bits}\right\rbrace
\end{equation} 
\begin{equation}
	\label{Eq. SequenceTranHD} 
	\ws{\Rk{\kappa}}{1}{K} = \left\lbrace \underbrace{0,\vf\cdots\vf ,0}_\text{First $\kappa$ bits}, \underbrace{ \wt{\Rk{\kappa}}{\kappa \vf + \vf 1}, 0, \wt{\Rk{\kappa}}{\kappa \vf + \vf 3}, 0,\vf \cdots \vf , 0, \wt{\Rk{\kappa}}{2L \vf + \vf \kappa \vf - \vf 1} }_\text{$L$ bits of information zero-padded with $(L-1)$ $0$s } 
	 \underbrace{0, \cdots, 0}_\text{Last $Q-\kappa$ bits}  \right\rbrace, 
\end{equation} 
\hrulefill
\end{figure*}

\subsection{Preliminaries} 
In the following, we consider communication between a transmitting node $\xstar \in \{S,\Rk{1},...,\Rk{\kappa}\}$ and a receiving node $\ystar \in \{\Rk{1},..., \Rk{\kappa}, D\}$, $\xstar \neq \ystar$, where $\xstar$ is the only transmitting node in the network, and review the corresponding error rate analysis as reported in \cite{NoelJ1}, \cite{NoelJ2}. These results are utilized in the analysis of the multi-hop network in Sections \ref{Sec.Two-hopNetPerAna}-\ref{Sec. SM-MH}. In the following, $A$ is the type of molecule released by node $\xstar$ and detected at node $\ystar$, i.e., we drop the subscript $f$ for clarity. 

The independent diffusion of molecules through the environment can be described by Fick's second law as \cite[Eq. (3)]{NoelJ1} 
\begin{equation}
	\label{Eq. Fick's Second Law} 
	\frac{\partial C_{A}(\vec{r},t)}{\partial t} = D_{A} \nabla^2 C_{A}(\vec{r},t),
\end{equation} 
where $D_{A}$ is the diffusion coefficient of the $A$ molecules in $\frac{\text{m}^2}{\text{s}}$. Assuming that node $\xstar$ is an impulsive point source, and emits $N_{A}$ molecules at the point defined by vector $\vec{r}_{\xstar}$ into an infinite environment at time $t = 0$, then the local concentration at the point defined by vector $\vec{r}$ and at time $t$ is given by \cite[Eq. (4)]{NoelJ1} 
\begin{equation}
	\label{Eq. Local Concentration} 
	C_{A}(\vec{r},t) = \frac{N_{A}}{(4 \pi D_{A} t)^{3/2}} \exp \left( -\frac{\vert \vec{r} - \vec{r}_{\xstar}\vert^2}{4D_{A}} \right).
\end{equation}

It is also shown in \cite{NoelJ1} that the number of molecules observed within the volume of node $\ystar$, $V_{\ystar}$, at time $t$ due to one emission of $N_{A}$ molecules at $\vec{r}_{\xstar}$ at $t=0$, $\Nob{}{\xstar}{\ystar}(t)$, can be accurately approximated as a Poisson random variable with time-varying mean given by
\begin{equation}
	\label{Eq. NobE} 
	\NobE{}{\xstar}{\ystar}(t) = C_{A}(\vec{r}_{\ystar},t)V_{\ystar},
\end{equation}
where $\vec{r}_{\ystar}$ is the vector from the origin to the center of node $\ystar$, and we used the uniform concentration assumption, i.e., we assumed that node $\ystar$ is a point observer or that the concentration throughout its volume is uniform and equal to that at its center. This assumption is accurate if node $\ystar$ is sufficiently far from node $\xstar$; see \cite{NoelPro1}. The probability of observing a given $A$ molecule, emitted by node $\xstar$ at time $t=0$, inside $V_{\ystar}$ at time $t$, $\Pob{}{\xstar}{\ystar}(t)$, is given by (\ref{Eq. NobE}) when setting $N_{A}=1$, i.e.,
\begin{equation}
	\label{Eq. Pob} 
	\Pob{}{\xstar}{\ystar}(t) = \frac{V_{\ystar}}{(4 \pi D_{A} t)^{3/2}} \exp \left( -\frac{\vert \vec{r}_{\ystar} -\vec{r}_{\xstar}\vert^2}{4D_{A}} \right). 
\end{equation} 

For detection, we adopt a family of receivers introduced in \cite{NoelJ2}, the so-called weighted sum detectors, where the receiving node takes $M \geq 1$ samples within a single bit interval, adds the individual samples with a certain weight assigned to each sample, and then compares the sum with a decision threshold. For simplicity, we assume equally spaced samples in time, and equal weights for all samples. The decision of the weighted sum in the $j$th bit interval is then given by \cite[Eq. (37)]{NoelJ2} 
\begin{equation}
	\label{Eq.Reception} 
	\wh{\ystar}{j} = \begin{cases} 
	1 &\mbox{if } \sum_{m=1}^{M} \Nob{}{\xstar}{\ystar}(t(j,m)) \geq \xi_{\ystar}, \\
	0 &\mbox{otherwise,} 
			\end{cases}
\end{equation} 
where $\xi_{\ystar}$ is the detection threshold of node $\ystar$. The sampling time of the $m$th sample in the $j$th bit interval is $t(j,m) = (j-1)T + t_{m}$, where $t_{m}=mt_{0}$ and $t_{0}$ is the time between two successive samples. $\Nob{}{\xstar}{\ystar}(t(j,m))$ is a Poisson random variable with mean $\NobE{}{\xstar}{\ystar}(t(j,m))$ for any individual sample. Thus, the sum of all samples in the $j$th bit interval, $\Nob{}{\xstar}{\ystar}[j]=\sum_{m=1}^{M} \Nob{}{\xstar}{\ystar}(t(j,m))$, is also a Poisson random variable whose mean is the sum of the means of the individual samples, i.e., $\NobE{}{\xstar}{\ystar}[j] = \sum_{m=1}^{M}\NobE{}{\xstar}{\ystar}(t(j,m))$. Due to the independent movement of molecules, node $\ystar$ observes molecules that were emitted by node $\xstar$ at the start of the current or any prior bit interval. As a result, the number of molecules observed within $V_{\ystar}$ in the $j$th bit interval due to the transmission of sequence $\ws{\xstar}{1}{j}$, $\Nob{}{\xstar}{\ystar}[j]$, is also a Poisson random variable with mean 
\begin{equation}
	\label{Eq. NobE Cumulative} 
	\NobE{}{\xstar}{\ystar}[j] = N_{A} \sum_{i=1}^{j} \wt{\xstar}{i} \sum_{m=1}^{M} \Pob{}{\xstar}{\ystar}((j-i)T+t_{m}).
\end{equation} 

Given $\ws{\xstar}{1}{j-1}$ and assuming that there is no \textit{a priori} knowledge about $\wt{\xstar}{j}$, the probability of error in the $j$th bit interval, $P_{e_{1}}[j|\ws{\xstar}{1}{j-1}]$, can be written as 
\begin{align} 
	\label{Eq. ExpErrorOneHop}
	P_{e_{1}}[j|\ws{\xstar}{1}{j-1}] &= P_{1}\pr(\Nob{}{\xstar}{\ystar}[j] < \xi_{\ystar} | \wt{\xstar}{j}=1,\ws{\xstar}{1}{j-1}) \nonumber \\
	& + P_{0}\pr(\Nob{}{\xstar}{\ystar}[j] \geq \xi_{\ystar} | \wt{\xstar}{j}=0,\ws{\xstar}{1}{j-1}),
\end{align} 
where the cumulative distribution function (CDF) of the weighted sum in the $j$th bit interval is given by \cite[Eq. (38)]{NoelJ2} 
\begin{align}
	\label{Eq. CDF} 
	\pr \left( \Nob{}{\xstar}{\ystar}[j] < \xi_{\ystar} | \ws{\xstar}{1}{j} \right) =&\; \exp (-\NobE{}{\xstar}{\ystar}[j]) \nonumber \\ 
      &\;\times \sum_{\omega=0}^{\xi_{\ystar}-1} \frac{\left( \NobE{}{\xstar}{\ystar}[j] \right)^{\omega}}{\omega!}.
\end{align}

The average error probability in the $j$th bit interval, $\overline{P}_{e_{1}}[j]$, is obtained by averaging $P_{e_{1}}[j|\ws{\xstar}{1}{j-1}]$ over all possible realizations of $\ws{\xstar}{1}{j-1}$, i.e., 
\begin{equation}
	\label{Eq. ExpErrorOneHopAvg} 
	\overline{P}_{e_{1}}[j] = \sum_{\ws{\xstar}{1}{j-1} \in \mathcal{W}} \pr(\ws{\xstar}{1}{j-1}) P_{e_{1}}[j|\ws{\xstar}{1}{j-1}],
\end{equation} 
where $\mathcal{W}$ is a set containing all realizations of $\ws{\xstar}{1}{j-1}$, and $\pr(\ws{\xstar}{1}{j-1})$ is the likelihood of the occurrence of $\ws{\xstar}{1}{j-1}$.

In the remainder of this paper, we denote the \emph{complete} received signal, i.e., due to molecules released by all nodes, at node $\ystar$ in the $j$th bit interval by $\NobTwo{}{\ystar}[j]$, and the received signal at node $\ystar$ in the $j$th bit interval that is originating from node $\xstar$ by $\Nob{}{\xstar}{\ystar}[j]$.
\section{MULTI-MOLECULE MULTI-HOP NETWORK}
\label{Sec.Two-hopNetPerAna} 
In this section, we evaluate the expected error probability of a multi-hop network if a unique type of molecule is used in each hop. We first derive a closed-form expression for the expected error probability of a two-hop network, as presented in \cite{Arman1}, and then generalize our analysis to the multi-hop case. Subsequently, we minimize the expected error probability of individual hops of the MM-MH network by finding the optimal number of released molecules and the optimal detection threshold in each hop.
\subsection{Two-hop Network} 
In the two-hop case, node $S$ emits type $A_{1}$ molecules, which have diffusion coefficient $D_{A_{1}}$ and can be detected by relay node $\Rk{1}$. The relay emits type $A_{2}$ molecules having diffusion coefficient $D_{A_{2}}$ for forwarding the detected message to node $D$, cf. Fig. \ref{Fig.SysMod}. Node $S$ and node $\Rk{1}$ release $N_{A_{1}}$ and $N_{A_{2}}$ molecules to transmit bit ``1'' at the beginning of a bit interval, respectively.

Since molecules of different types do not interfere with each other, we only consider full-duplex relaying in this case. The nodes communicate as follows. At the beginning of the $j$th bit interval, node $S$ transmits information bit $\wt{S}{j}$, and node $\Rk{1}$ concurrently transmits the information bit detected in the previous bit interval, $\wt{\Rk{1}}{j} = \wh{\Rk{1}}{j-1}$. At the end of the $j$th bit interval, node $\Rk{1}$ and node $D$ make decisions on the respective received signals. 

In the two-hop communication link, for binary modulation, an error occurs if the detection is erroneous in either the first hop or the second hop. Given $\wt{S}{j}$, an error occurs in the $(j+1)$th bit interval if $\wh{\Rk{1}}{j} \neq \wt{S}{j}$ and $\wh{D}{j+1} = \wt{\Rk{1}}{j+1}$, or if $\wh{\Rk{1}}{j} = \wt{S}{j}$ and $\wh{D}{j+1} \neq \wt{\Rk{1}}{j+1}$. Thus, the error probability of the $j$th bit can be written as 
\begin{align}
	\label{Eq. ErrorExpCurBit} 
	P_{e_{2}}[j|\wt{S}{j}] \vf &= \vf \pr( \wt{S}{j} \vf \neq \vf \wh{\Rk{1}}{j})  \pr(\wt{\Rk{1}}{j \vf + \vf 1} \vf = \vf \wh{D}{j \vf + \vf 1}) \nonumber \\
	& + \vf \pr( \wt{S}{j} \vf = \vf \wh{\Rk{1}}{j})  \pr(\wt{\Rk{1}}{j \vf + \vf 1} \vf \neq \vf \wh{D}{j \vf + \vf 1}).
\end{align}        
Let us assume that $\ws{S}{1}{j-1}$ is given, then the error probability of the $j$th bit when $\wt{S}{j}=1$ and $\wt{S}{j}=0$ can be written as 
\begin{align}
	\label{Eq. ErrorExpSeqW1} 
	  P_{e_{2}}[j|\wt{S}{j} \vf = \vf 1, \vf \ws{S}{1}{j-1}] \vf &= \vf \pr (\NobTwo{1}{\Rk{1}}[j] \vf < \vf \xiN{\Rk{1}} | \wt{S}{j} \vf = \vf 1, \vf \ws{S}{1}{j-1}) \nonumber \\
	 & \hspace{-11.5mm} \times \pr(\NobTwo{2}{D}[j \vf + \vf 1] \vf < \vf \xiN{D} | \wt{\Rk{1}}{j \vf + \vf 1} \vf = \vf 0, \wsh{\Rk{1}}{1}{j-1}) \nonumber \\ 
	& + \vf \pr (\NobTwo{1}{\Rk{1}}[j] \vf \geq \vf \xiN{\Rk{1}} | \wt{S}{j} \vf = \vf 1, \vf \ws{S}{1}{j-1}) \nonumber \\  
	 & \hspace{-11.5mm} \times \pr(\NobTwo{2}{D}[j \vf + \vf 1] \vf < \vf \xiN{D} | \wt{\Rk{1}}{j \vf + \vf 1} \vf = \vf 1, \vf \wsh{\Rk{1}}{1}{j-1}),
\end{align} 
and 
\begin{align}
	\label{Eq. ErrorExpSeqW0} 
	  P_{e_{2}}[j|\wt{S}{j} \vf = \vf 0, \vf \ws{S}{1}{j-1}] \vf &= \vf \pr (\NobTwo{1}{\Rk{1}}[j] \vf \geq \vf \xiN{\Rk{1}} | \wt{S}{j} \vf = \vf 0, \vf \ws{S}{1}{j-1}) \nonumber \\
	 & \hspace{-11.5mm} \times \pr(\NobTwo{2}{D}[j \vf + \vf 1] \vf \geq \vf \xiN{D} | \wt{\Rk{1}}{j \vf + \vf 1} \vf = \vf 1, \wsh{\Rk{1}}{1}{j-1}) \nonumber \\ 
	  & + \vf \pr (\NobTwo{1}{\Rk{1}}[j] \vf < \vf \xiN{\Rk{1}} | \wt{S}{j} \vf = \vf 0, \vf \ws{S}{1}{j-1}) \nonumber \\
	 & \hspace{-11.5mm} \times \pr(\NobTwo{2}{D}[j \vf + \vf 1] \vf \geq \vf \xiN{D} | \wt{\Rk{1}}{j \vf + \vf 1} \vf = \vf 0, \vf \wsh{\Rk{1}}{1}{j-1}),
\end{align} 
respectively, where $\NobTwo{1}{\Rk{1}}[j] = \Nob{1}{S}{\Rk{1}}[j]$, $\NobTwo{2}{D}[j] = \Nob{2}{\Rk{1}}{D}[j]$, and the involved probabilities can be obtained based on $(\ref{Eq. CDF})$. If we do not have knowledge about $\wt{S}{j}$, then the expected error probability is given by 
\begin{align}
	\label{Eq. ErrorExpSeq} 
	  P_{e_{2}}[j|\ws{S}{1}{j-1}] =&\; P_{1}P_{e_{2}}[j|\wt{S}{j} = 1, \ws{S}{1}{j-1}] \nonumber \\
	  &\;+ P_{0}P_{e_{2}}[j|\wt{S}{j} = 0, \ws{S}{1}{j-1}].
\end{align} 
For a given $\ws{S}{1}{j-1}$, there are $2^{(j-1)}$ different possible realizations of $\wsh{\Rk{1}}{1}{j-1}$. However, in (\ref{Eq. ErrorExpSeqW1}) and (\ref{Eq. ErrorExpSeqW0}), to keep the complexity of evaluation low, we consider only one realization of $\wsh{\Rk{1}}{1}{j-1}$ which leads to an approximation. In particular, this realization of $\wsh{\Rk{1}}{1}{j-1}$ is obtained via a biased coin toss. To this end, we model the detected bits in $\wsh{\Rk{1}}{1}{j-1}$ , i.e., $\wh{\Rk{1}}{i}, i \in \{1,2,...,j-1 \}$, as $\wh{\Rk{1}}{i} = |\lambda - \wt{S}{i} |$, where $\lambda \in \{0,1 \}$ is the outcome of the coin toss with $\pr(\lambda = 1) = P_{e_{1}}[i|\ws{S}{1}{i-1}]$ and $\pr(\lambda = 0) = 1-P_{e_{1}}[i|\ws{S}{1}{i-1}]$. Our simulation results in Section \ref{Sec.NumRes} confirm the accuracy of this approximation.

\subsection{Multi-hop Network}
\begin{figure}[t]
	\centering
	\includegraphics[scale = 0.55]{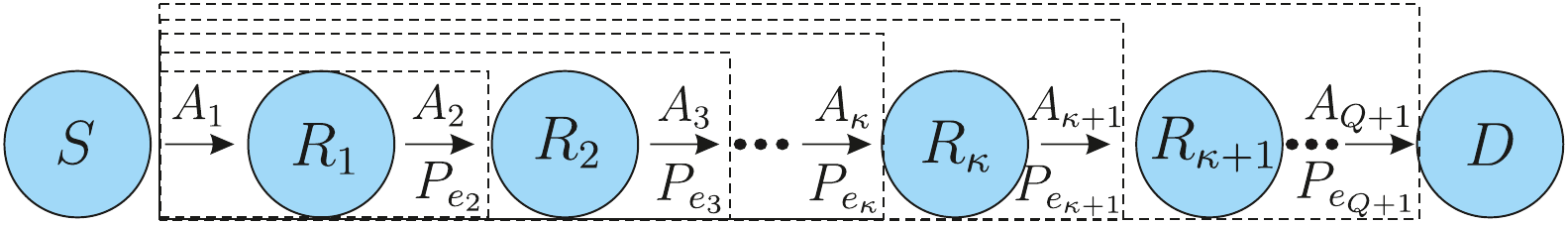}
	\caption{Illustration of a MM-MH network, where $P_{e_{\kappa}}$ denotes the expected error probability after the first $\kappa$ hops of the network.}
	\label{Fig.ErrorRate}
\end{figure}
We now extend our analysis to the multi-hop case. Relay node $R_{\kappa}, \kappa \in \{1,2,...,Q  \}$, detects type $A_{\kappa}$ messenger molecules, with diffusion coefficient $D_{A_{\kappa}}$, that are released by node $\Rk{\kappa - 1}$, and emits messenger molecules of type $A_{\kappa+1}$, with diffusion coefficient $D_{A_{\kappa+1}}$, to forward the detected message to node $\Rk{\kappa + 1}$. We assume that node $D$ can only detect the type of molecule emitted by the last relay, $A_{Q+1}$, cf. Fig. \ref{Fig.SysMod}. Thus, there is no direct communication path between node $S$ and node $D$. 

In the following, we propose a recursive algorithm to evaluate the expected error probability of the multi-hop network, where in the $\kappa$th iteration of the algorithm, we evaluate the expected error probability of the first $\kappa + 1$ hops of the network. To this end, we define $\pe{\kappa}[j|\wt{S}{j} = \{0,1 \},\ws{S}{1}{j-1}]$ as the conditional expected error probability of the first $\kappa$ hops in the $j$th bit interval, i.e., the probability that node $S$ sends $\wt{S}{j} = \{0,1 \}$ and $\wh{\Rk{\kappa}}{j+\kappa-1} \neq \wt{S}{j}$ is detected at node $\Rk{\kappa}$. Let us assume that $\pe{\kappa}[j|\wt{S}{j} = \{0,1 \}, \ws{S}{1}{j-1}]$ is known. In order to evaluate $\pe{\kappa + 1}[j|\wt{S}{j} = \{0,1 \},\ws{S}{1}{j-1}]$, we form a virtual two-hop link where the first $\kappa$ hops of the multi-hop network are modeled as the first hop of the virtual two-hop link, cf. Fig \ref{Fig.ErrorRate}. Thus, the expected error probability of the first $(\kappa+1)$ hops of the multi-hop network when $\wt{S}{j}=1$ and $\wt{S}{j}=0$ can be written as 
\begin{align}
	\label{Eq. ErrorExpSeqmulti-hopW1} 
	  P_{e_{\kappa + 1}}[j|\wt{S}{j}=1,\ws{S}{1}{j-1}] \vf &= \vf \pe{\kappa}[j | \wt{S}{j} \vf = \vf 1, \ws{S}{1}{j-1}] \nonumber \\ 
	 & \hspace{-29mm} \times \pr(\NobTwo{\kappa + 1}{\Rk{\kappa + 1}}[j \vf + \vf \kappa] \vf < \vf \xiN{\Rk{\kappa + 1}} | \wt{\Rk{\kappa}}{j \vf + \vf \kappa} \vf = \vf 0, \wsh{\Rk{\kappa}}{1}{j+\kappa-2}) \nonumber \\ 
	& + \vf (1 \vf - \vf \pe{\kappa}[j| \wt{S}{j} \vf = \vf 1, \ws{S}{1}{j-1}]) \nonumber \\  
	 & \hspace{-29mm} \times \pr(\NobTwo{\kappa + 1}{\Rk{\kappa + 1}}[j \vf + \vf \kappa] \vf < \vf \xiN{\Rk{\kappa + 1}} | \wt{\Rk{\kappa}}{j \vf + \vf \kappa} \vf = \vf 1, \wsh{\Rk{\kappa}}{1}{j + \kappa -2}) 
\end{align} 
and 
\begin{align}
	\label{Eq. ErrorExpSeqmulti-hopW0} 
	  P_{e_{\kappa + 1}}[j|\wt{S}{j}=0,\ws{S}{1}{j-1}] \vf &= \vf \pe{\kappa}[j | \wt{S}{j} \vf = \vf 0, \ws{S}{1}{j-1}] \nonumber \\
	 & \hspace{-29mm} \times \pr(\NobTwo{\kappa + 1}{\Rk{\kappa + 1}}[j \vf + \vf \kappa] \vf \geq \vf \xiN{\Rk{\kappa + 1}} | \wt{\Rk{\kappa}}{j \vf + \vf \kappa} \vf = \vf 1, \wsh{\Rk{\kappa}}{1}{j+\kappa -2}) \nonumber \\ 
	& + \vf (1 \vf - \vf \pe{\kappa}[j | \wt{S}{j} \vf = \vf 0, \ws{S}{1}{j-1}]) \nonumber \\
	 & \hspace{-29mm} \times \pr(\NobTwo{\kappa+1}{\Rk{\kappa+1}}[j \vf + \vf \kappa] \vf \geq \vf \xiN{\Rk{\kappa+1}} | \wt{\Rk{\kappa}}{j \vf + \vf \kappa} \vf = \vf 0, \wsh{\Rk{\kappa}}{1}{j+\kappa-2}),
\end{align}
respectively, where $\NobTwo{\kappa+1}{\Rk{\kappa+1}}[j] = \Nob{\kappa+1}{\Rk{\kappa}}{\Rk{\kappa+1}}[j]$. If we do not have \emph{a priori} knowledge about $\wt{S}{j}$, then $P_{e_{\kappa + 1}}[j|\ws{S}{1}{j-1}]$ is given by  
\begin{align}
	\label{Eq. ErrorExpSeqmulti-hop} 
	  P_{e_{\kappa + 1}}[j|\ws{S}{1}{j-1}]  =&\;  P_{1} P_{e_{\kappa + 1}}[j|\wt{S}{j}=1,\ws{S}{1}{j-1}] \nonumber \\
	  &\;+ P_{0}P_{e_{\kappa + 1}}[j|\wt{S}{j}=0,\ws{S}{1}{j-1}].
\end{align}   
Similar to the two-hop case, the previously detected bits in $\wsh{\Rk{\kappa}}{1}{j+\kappa -2}$, i.e., $\wh{\Rk{\kappa}}{i}, i \leq j + \kappa -2$, are modeled as $\wh{\Rk{\kappa}}{i} = |\lambda - \wt{S}{i}|$, where $\lambda \in \{ 0,1 \}$ is the outcome of a coin toss with $\pr(\lambda=1)= P_{e_{\kappa}}[i|\ws{S}{1}{i-1}] $ and $\pr(\lambda =0)=1-\pr(\lambda=1)$. Given $\ws{S}{1}{j-1}$, the proposed algorithm for evaluation of the expected error probability of an MM-MH network consisting of $Q > 1$ relays, $P_{e_{Q + 1}}[j|\ws{S}{1}{j-1}]$, is summarized in Algorithm \ref{Alg. ErrorExpSeqmulti-hop}. 
\begin{figure}[!h] 
 \removelatexerror
  \begin{algorithm}[H] 
  \label{Alg. ErrorExpSeqmulti-hop}
   \caption{Evaluation of \( P_{e_{Q+1}}{[j|\ws{S}{1}{j-1}]} \)}
   Initialization: Given $\ws{S}{1}{j-1}$  
   evaluate $P_{e_{2}}[j|\wt{S}{j} \vf = \vf 1, \ws{S}{1}{j-1}]$ and $P_{e_{2}}[j|\wt{S}{j} \vf = \vf 0, \ws{S}{1}{j-1}]$ via (\ref{Eq. ErrorExpSeqW1}) and (\ref{Eq. ErrorExpSeqW0}), respectively. \\
   \For{$\kappa = 2:Q$}
   {
      Evaluate $P_{e_{\kappa + 1}}[j|\wt{S}{j}=1,\ws{S}{1}{j-1}]$ via (\ref{Eq. ErrorExpSeqmulti-hopW1}), given $P_{e_{\kappa}}[j|\wt{S}{j}=1,\ws{S}{1}{j-1}]$ calculated in the previous iteration. \\  
      Evaluate $P_{e_{\kappa + 1}}[j|\wt{S}{j}=0,\ws{S}{1}{j-1}]$ via (\ref{Eq. ErrorExpSeqmulti-hopW0}), given $P_{e_{\kappa}}[j|\wt{S}{j}=0,\ws{S}{1}{j-1}]$ calculated in the previous iteration. \\ 
   } 
   Evaluate $P_{e_{Q + 1}}[j|\ws{S}{1}{j-1}]$ via (\ref{Eq. ErrorExpSeqmulti-hop}).
  \end{algorithm}
\end{figure}
\subsection{Parameter Optimization} 
The performance of a MM-MH network can be improved by minimizing the error probability in each hop. To this end, in this section, we consider a single communication link between a transmitting node $\xstar \in \{S,\Rk{1},...,\Rk{\kappa}\}$ and the corresponding receiving node $\ystar \in \{\Rk{1},..., \Rk{\kappa}, D\}$, $\ystar = \xstar +1$, and derive closed-form expressions for both the optimal detection threshold, $\xi_{opt}$, of node $\ystar$, and the optimal number of molecules, $N_{A_{opt}}$, released by node $\xstar$ for minimization of  the expected error probability of this single link.

\subsubsection{Optimal Number of Released Molecules}
Eq. (\ref{Eq. NobE Cumulative}) specifies the expected number of molecules observed within $V_{\ystar}$ due to the transmission of $\ws{\xstar}{1}{j}$. This equation can be re-written as 
\begin{align}
	\label{Eq. NobE Cumulative Two-term} 
	\NobE{}{\xstar}{\ystar}[j] =&\; N_{A} \sum_{i=1}^{j-1} \wt{\xstar}{i} \sum_{m=1}^{M} \Pob{}{\xstar}{\ystar}((j-i)T+t_{m}) \nonumber \\
	&\;+ N_{A} \wt{\xstar}{j} \sum_{m=1}^{M} \Pob{}{\xstar}{\ystar}(t_{m}),
\end{align} 
where the first term is the expected number of observed molecules due to the transmission of all previous bits (ISI), $\wt{\xstar}{i}, i<j$, and the second term is the expected number of molecules observed due to the transmission of the most recent bit, $\wt{\xstar}{j}$. From (\ref{Eq. NobE Cumulative Two-term}), and for a given $\xi_{\ystar}$, we observe that decreasing $N_A$ reduces the effect of ISI, and increases the probability of miss detection, i.e., $\pr(\wh{\ystar}{j} \neq \wt{\xstar}{j}| \wt{\xstar}{j}=1)$. On the other hand, increasing $N_A$ enhances the effect of ISI, and increases the probability of false alarm, i.e., $\pr(\wh{\ystar}{j} \neq \wt{\xstar}{j} | \wt{\xstar}{j}=0)$. Thus, the expected error probability in (\ref{Eq. ExpErrorOneHopAvg}) can be minimized by optimizing the number of released molecules $N_A$. In (\ref{Eq. ExpErrorOneHopAvg}), any realization of $\ws{\xstar}{1}{j-1}$ is independent of all other realizations. As a result, we can minimize (\ref{Eq. ExpErrorOneHop}) for a given $\ws{\xstar}{1}{j-1}$. $\Nob{}{\xstar}{\ystar}[j]$ is a Poisson random variable that has a \emph{discrete} distribution, which complicates the optimization. However, it is shown in \cite{NoelJ2} that the CDF of a Poisson random variable $X$ with mean $\rho$ can be accurately approximated by a \emph{continuous} regularized incomplete gamma function $Q(\cdot,\cdot)$, $\pr(X < s) = Q( \lceil s \rceil,\rho)= \Gamma(\lceil s \rceil, \rho)/ \Gamma(\lceil s \rceil)$ for $s > 0$, where $\Gamma(s, \rho)$ is the incomplete Gamma function given by \cite[Eq. (6.5.3)]{abramowitz} 
\begin{equation} 
	\label{Eq. Incomplete Gamma Function}
	\Gamma(s,\rho) = \int_{\rho}^{\infty} e^{-t}t^{s-1} dt.
\end{equation} 
 
The Gamma function, $\Gamma(s)$, is a special case of (\ref{Eq. Incomplete Gamma Function}) with $\rho=0$. Approximating the discrete CDF of a Poisson random variable by the continuous regularized incomplete Gamma function leads to an approximation of the optimal number of released molecules.
 
\textit{Proposition 1:} Given $\ws{\xstar}{1}{j-1}$, the optimal number of molecules released by node $\xstar$ in the beginning of the $j$th bit interval, $N_{A_{opt}}[j]$, that minimizes the expected error probability of a single link can be approximated as 
\begin{equation}
	\label{Eq. Optimum Na}
	N_{A_{opt}}[j]= \bigg\lfloor \frac{\ln \left( \frac{P_{1}}{P_{0}} \right) + \xi_{\ystar} \ln \left( \frac{m_{1}[j] }{ m_{0}[j] } \right)}{\sum_{m=1}^{M} \Pob{}{\xstar}{\ystar}(t_{m})} \bigg\rceil,
\end{equation} 
where $\ln(\cdot)$ is the natural logarithm, $\lfloor\cdot\rceil$ is the nearest integer, $m_{x}[j], x \in \{ 0,1 \}$, is given by  
\begin{equation}
	\label{Eq. Cx} 
	m_{x}[j] = \NobE{}{\xstar}{\ystar}[j] \big|_{ \wt{\xstar}{j}=x, \ws{\xstar}{1}{j-1}},
\end{equation} 
and $\Pob{}{\xstar}{\ystar}(\cdot)$ is given in (\ref{Eq. Pob}). $m_{x}[j]$ is the conditional mean of Poisson random variable $\Nob{}{\xstar}{\ystar}[j]$ in the $j$th bit interval when the most recent information bit transmitted by node $\xstar$ is $\wt{\xstar}{j}=x$, and can be evaluated based on (\ref{Eq. NobE Cumulative Two-term}). When $P_{1}=P_{0}$, (\ref{Eq. Optimum Na}) simplifies to 
\begin{equation}
	\label{Eq. Optimum Na p1=p0} 
	N_{A_{opt}}[j]= \bigg\lfloor \frac{ \xi_{\ystar} \ln \left( \frac{ m_{1}[j] }{m_{0}[j]} \right)}{\sum_{m=1}^{M} \Pob{}{\xstar}{\ystar}(t_{m})} \bigg\rceil.
\end{equation}
\begin{IEEEproof}
The partial derivative of the incomplete Gamma function with respect to its second elementary variable is given by \cite[Eq. (6.5.25)]{abramowitz} $\partial \Gamma(s,\rho) / \partial \rho = -e^{-\rho} \rho^{s-1}.$ By taking the derivative of (\ref{Eq. ExpErrorOneHop}) with respect to continuous $N_{A}$, solving the resulting equation for $N_{A}$, and rounding the solution to the nearest integer number, the optimal number of molecules released by node $\xstar$ can be written as (\ref{Eq. Optimum Na}). 
\end{IEEEproof}  
\textit{Remark 1:} Eq. (\ref{Eq. Optimum Na p1=p0}) provides insight for the selection of $N_{A_{opt}}[j]$ based on the other system parameters. In particular, $N_{A_{opt}}[j]$ scales linearly with $\xi$, i.e., if the detection threshold at node $\ystar$ is doubled, node $\xstar$ should transmit twice as many molecules. Intuitively, if the ISI increases, e.g., because the bit interval duration is decreased, then we would expect that node $\xstar$ should release fewer molecules. In (\ref{Eq. Optimum Na p1=p0}), when the ISI increases, $\frac{ m_{1}[j] }{m_{0}[j]}$ approaches one, and, as a result, $N_{A_{opt}}[j]$ approaches zero. Thus, (\ref{Eq. Optimum Na}) and (\ref{Eq. Optimum Na p1=p0}) can be used to select the optimal number of released molecules for a specific modulation bit interval duration, and/or a certain distance between node $\xstar$ and node $\ystar$. The average optimal number of released molecules, $\overline{N}_{A_{opt}}$, can be evaluated by taking the average of $N_{A_{opt}}[j]$ over all realizations of $\ws{\xstar}{1}{j-1}$, and averaging over all bit intervals.

\subsubsection{Optimal Detection Threshold} 
We now consider optimizing the detection threshold at the receiver side. We assume that node $\xstar$ releases a fixed number of molecules, $N_{A}$, for bit ``1'', and no molecules for bit ``0''. The simulation results in \cite{NoelJ1} reveal that the performance of the receiver depends on the value of the chosen detection threshold $\xi$. Increasing $\xi$ enhances the probability of miss detection. However, decreasing $\xi$ increases the probability of false alarm. Thus, we are interested in the optimal detection threshold, $\xi_{opt}$, that minimizes (\ref{Eq. ExpErrorOneHopAvg}).

\textit{Proposition 2:} Given $\ws{\xstar}{1}{j-1}$, the optimal detection threshold at the receiver in the $j$th bit interval, that minimizes the expected error probability of a single link, can be approximated as 
\begin{equation}
	\label{Eq. Optimum xi}
	\xi_{opt}[j] = \bigg\lfloor \frac{ \ln \left( \frac{P_{0}}{P_{1}} \right) + N_{A} \sum_{m=1}^{M} \Pob{}{\xstar}{\ystar}(t_{m})}{ \ln \big( \frac{m_{1}[j]}{m_{0}[j]} \big)} \bigg\rceil.
\end{equation} 
 
For $P_{1}=P_{0}$, (\ref{Eq. Optimum xi}) simplifies to 
\begin{equation}
	\label{Eq. Eq. Optimum xi P1=P0} 
	\xi_{opt}[j] = \bigg\lfloor\frac{ N_{A} \sum_{m=1}^{M} \Pob{}{\xstar}{\ystar}(t_{m}) }{ \ln \big( \frac{m_{1}[j]}{m_{0}[j]} \big)}\bigg\rceil.
\end{equation} 

\begin{IEEEproof} 
To find (\ref{Eq. Optimum xi}), we have to take the partial derivative of (\ref{Eq. ExpErrorOneHop}) with respect to $\xi_{\ystar}$, which, in turn, requires the partial derivative of (\ref{Eq. CDF}) with respect to its first elementary variable, $\xi_{\ystar}$, if we approximate (\ref{Eq. CDF}) with $Q(\lceil \xi_{\ystar} \rceil, \NobE{}{\xstar}{\ystar})$. However, since there is no closed-form expression for this partial derivative, we approximate the factorial term in (\ref{Eq. CDF}) with the Stirling formula, $\omega ! \simeq (2 \pi \omega)^{1/2} (\omega / e)^{\omega}$. Thus, it can be shown that the discrete CDF of Poisson random variable $\Nob{}{\xstar}{\ystar}[j]$ with mean $\NobE{}{\xstar}{\ystar}[j]$, Eq. (\ref{Eq. CDF}), can be approximated with a continuous CDF as 
\begin{align}
	\label{Eq. pdf Poisson factorial approximation} 
	\pr(\Nob{}{\xstar}{\ystar}[j] < \xi_{\ystar} | \ws{\xstar}{1}{j}) &\simeq \nonumber \\
	& \hspace{-15mm} \int_{0}^{\xi_{\ystar}} \frac{e^{(\omega - \NobE{}{\xstar}{\ystar}[j])} \left(\frac{\NobE{}{\xstar}{\ystar}[j]}{\omega}\right)^{\omega + 1/2} }{\sqrt[]{2 \pi \NobE{}{\xstar}{\ystar}[j]}} d \omega,
\end{align} 

Using (\ref{Eq. pdf Poisson factorial approximation}) as the continuous approximation of the CDF of a discrete Poisson random variable, taking the derivative of (\ref{Eq. ExpErrorOneHop}) with respect to continuous $\xi_{\ystar}$, solving the resulting equation for $\xi_{\ystar}$, and rounding the solution to the nearest integer value, the optimal detection threshold at node $\ystar$ can be expressed as in (\ref{Eq. Optimum xi}).  
\end{IEEEproof}  

\textit{Remark 2:} Eqs. (\ref{Eq. Optimum xi}) and (\ref{Eq. Eq. Optimum xi P1=P0}) can be used to select the optimal detection threshold at the receiver side for different parameters of the system such as the modulation bit interval, the distance between node $\xstar$ and node $\ystar$, the number of samples per bit interval, $M$, and the number of released molecules by node $\xstar$, $N_A$. The average optimal detection threshold, $\overline{\xi}_{opt}$, can be evaluated by averaging (\ref{Eq. Optimum xi}) over all bit intervals and realizations of $\ws{\xstar}{1}{j-1}$. 

\textit{Remark 3:} Eqs. (\ref{Eq. Optimum Na}) and (\ref{Eq. Optimum xi}) could be solved iteratively to find the jointly optimal $\xi_{opt}[j]$ and $N_{A_{opt}}[j]$. However, we do not consider the joint optimization of $\xi_{opt}[j]$ and $N_{A_{opt}}[j]$, since solving (\ref{Eq. Optimum xi}) for $N_A$ leads to (\ref{Eq. Optimum Na}). Hence, as far as the performance is concerned, optimizing $\xi_{\ystar}$ is equivalent to optimizing $N_A$. Thus, the parameters to be optimized can be chosen for implementation convenience.
\section{TWO-MOLECULE MULTI-HOP NETWORK} 
\label{Sec. TM-MH}
In this section, we consider 2M-MH, where we assume that only two different types of molecules are available as information carriers, namely type $A_{1}$ and type $A_{2}$ molecules. Since node $\Rk{\kappa}$ uses two different molecules for reception and transmission, we focus on full-duplex transmission. In the $j$th bit interval, relay node $\Rk{\kappa}$ detects the information bit transmitted by node $\Rk{\kappa -1}$, and sends the information bit detected in the previous bit interval to node $\Rk{\kappa + 1}$. Set $\mathcal{R_E} \in \{ \Rk{2i} \}, 0 \leq i \leq \lfloor \frac{Q+1}{2} \rfloor $, includes all relays with even index, which release and detect type $A_1$ and $A_2$ molecules, respectively. Analogously, set $\mathcal{R_O} \in \{ \Rk{2i + 1} \}, 0 \leq i \leq \lfloor \frac{Q-1}{2} \rfloor$, includes all relays with odd index, which release and detect type $A_2$ and $A_1$ molecules, respectively.
 
In this scenario, $A_1$ molecules released by the relays belonging to set $\mathcal{R_E}$ may interfere with each other at relay $\Rk{\kappa} \in \mathcal{R_O}$, and $\Rk{\kappa}$ cannot differentiate between $A_1$ molecules released by $\Rk{\kappa -1}$ and interfering $A_1$ molecules released by the other relays in set $\mathcal{R_E} \backslash \Rk{\kappa-1}$. We refer to this effect, which is caused by the random walks of the molecules, as \emph{backward-ISI} when the interfering molecules were released by relays in future hops, i.e., by relays $\Rk{2i} \in \mathcal{R_E}, 2i > \kappa$, and as \emph{forward-ISI} when the interfering molecules were released by relays in previous hops, i.e., relays $\Rk{2i} \in \mathcal{R_E}, 2i < \kappa$. The same effects are present for the $A_2$ molecules released by the relays belonging to the set $\mathcal{R_O}$ at relay $\Rk{\kappa} \in \mathcal{R_E}$.

Let us consider a short example to clarify the occurrence of backward-ISI and forward-ISI. \textit{Example 1:} Let us assume that $\Rk{\kappa-2}$, $R_{\kappa-1}$, $R_{\kappa}$, and $R_{\kappa+1}, \kappa \neq \{1,Q\}$, are four consecutive relays that release type $A_2$, $A_1$, $A_2$, and $A_1$ molecules, respectively. Furthermore, assume that the information sequences transmitted by $R_{\kappa-2}$ and $\Rk{\kappa-1}$ are ``1000'' and ``0100'', respectively, i.e., $\wt{\Rk{\kappa-2}}{j} = \wt{\Rk{\kappa-1}}{j+1} = 1$, $\wt{\Rk{\kappa-2}}{j+1} = \wt{\Rk{\kappa-2}}{j+2} = \wt{\Rk{\kappa-2}}{j+3} = \wt{\Rk{\kappa-1}}{j} = \wt{\Rk{\kappa-1}}{j+2} = \wt{\Rk{\kappa-1}}{j+3} = 0$, $\wt{\Rk{\kappa-2}}{i}=0, i<j$, and $\wt{\Rk{\kappa-1}}{i}=0, i<j+1$. Due to the broadcast nature of the molecular channel, some of the $A_2$ molecules released by $\Rk{\kappa-2}$ at the beginning of the $j$th bit interval will be observed at $\Rk{\kappa+1}$ during the current and subsequent bit intervals at the time of sampling for detection of $\wh{\Rk{\kappa+1}}{i}, i \in \{j,j+1,j+2,j+3\}$. This causes forward-ISI and may lead to an erroneous decision for $\wh{\Rk{\kappa+1}}{i}$. Let us assume that no error occurs in the transmission of $\wt{\Rk{\kappa-1}}{j+1}$ to relay $\Rk{\kappa+1}$, i.e.,$\wh{\Rk{\kappa+1}}{j+2}=1$. Thus, in the beginning of the $(j+3)$th bit interval, $\Rk{\kappa+1}$ releases $N_{A_{1}}$ $A_{1}$ molecules to forward the message to $\Rk{\kappa+2}$, i.e., $\wt{\Rk{\kappa+1}}{j+3}=1$. Due to the broadcast nature of the molecular channel, some of the $A_{1}$ molecules released by $\Rk{\kappa+1}$ will be observed within $V_{R_{\kappa}}$ at the time of sampling for detection of $\wt{\Rk{\kappa-1}}{j+3}=0$. This causes backward-ISI and may lead to an erroneous decision for $\wh{\Rk{\kappa}}{j+3}$.
\subsection{Backward-ISI Mitigation} 
\label{SubSec.Backward-ISI}
In the following, we propose an algorithm to mitigate backward-ISI. In this algorithm, the relay node $\Rk{\kappa}, \kappa \neq Q$, adjusts its decision threshold in the $j$th bit interval based on its detected information bits in the previous $j-2$ bit intervals. The adaptive decision threshold of the relay $\Rk{\kappa}$ in the $j$th bit interval, $\xi_{\Rk{\kappa}}^{BI}[j]$, consists of two parts. The first part is a fixed threshold, $\xi$, and the second part, $\xi_{Exp}^{BI}[j]$, changes adaptively based on the number of molecules \emph{expected} within $V_{\Rk{\kappa}}$ due to the emissions of relay node $\Rk{\kappa+1}$, i.e., 
\begin{equation}
	\label{Eq.AdpThreshBI} 
	\xi_{\Rk{\kappa}}^{BI}[j] = \xi + \xi_{Exp}^{BI}[j].
\end{equation}
              
To optimize $\xi_{Exp}^{BI}[j]$, we require the probability of observing a given molecule transmitted by node $\Rk{\kappa+1}$ at time $t=0$ within $V_{\Rk{\kappa}}$ at time $t$. This probability can be evaluated via (\ref{Eq. Pob}) after substituting $\vec{r}_{\ystar}$ and $\vec{r}_{\xstar}$ with $\vec{r}_{\Rk{\kappa}}$ and $\vec{r}_{\Rk{\kappa+1}}$, respectively. Thus, given $\wsh{\Rk{\kappa}}{1}{j-2}$ and assuming that no error occurs in transmission of this sequence to $\Rk{\kappa+1}$, the expected number of molecules observed within $V_{\Rk{\kappa}}$ in the $j$th bit interval due to the transmission of $\wsh{\Rk{\kappa}}{1}{j-2}$ by $\Rk{\kappa+1}$ can be written as 
\begin{align} 
	\label{Eq. ExpBIOb} 
	\NobE{}{\Rk{\kappa+1}}{\Rk{\kappa}}[j] &= N_A \sum_{i=1}^{j-2} \wh{\Rk{\kappa}}{i} \nonumber \\
	& \times \sum_{m=1}^{M} \Pob{}{\Rk{\kappa+1}}{\Rk{\kappa}}((j-i)T+t_{m}), 
\end{align} 
where $A=A_1$ if $\Rk{\kappa} \in \mathcal{R_O}$ and $A=A_2$ if $\Rk{\kappa} \in \mathcal{R_E}$. Hence, the varying part of the adaptive decision threshold of node $\Rk{\kappa}$ is chosen as $\xi_{Exp}^{BI}[j] = \NobE{}{\Rk{\kappa+1}}{\Rk{\kappa}}[j]$. 

\textit{Remark 4:} In a multi-hop network consisting of $Q$ relays, the decision thresholds of relay node $\Rk{Q}$ and node $D$ ($\Rk{Q+1}$) are fixed to $\xi$, since backward-ISI does not occur at node $\Rk{Q}$ and node $D$. From this, we can also conclude that the smallest multi-hop network in which backward-ISI occurs is a three-hop network.

\textit{Remark 5:} In our proposed algorithm for mitigation of backward-ISI, we exploit the knowledge of the sequence transmitted by relay $\Rk{\kappa+1}$, $\ws{\Rk{\kappa+1}}{1}{j}$, which is obtained from the detected sequence at relay $\Rk{\kappa}$, $\wsh{\Rk{\kappa}}{1}{j-2}$, and the assumption that no error occurs in the transmission of $\wsh{\Rk{\kappa}}{1}{j-2}$ to $\Rk{\kappa+1}$. However, since the sequences transmitted by the relay nodes in hops $ i < \kappa$ are not known at $\Rk{\kappa}$, the adaptation of the decision threshold cannot be applied for mitigation of forward-ISI, and as a result, forward-ISI is the performance bottleneck of 2M-MH.       
\subsection{Performance Analysis}
For evaluation of the expected error probability of the considered network via (\ref{Eq. ErrorExpSeqmulti-hop}), we require the complete received signal at node $\Rk{\kappa+1}$. When $\kappa+1$ is odd, the complete received signal in the $j$th bit interval at node $\Rk{\kappa+1}$, $\NobTwo{1}{\Rk{\kappa+1}}[j]$, is the sum of all received signals transmitted by the relays in the set $\mathcal{R_E}$, i.e., 
\begin{equation}
\label{Eq. 2M-MHCompeleteSignalEven}
	\NobTwo{1}{\Rk{\kappa+1}}[j] = \sum_{i = 0}^{\lfloor \frac{Q+1}{2} \rfloor} \Nob{1}{\Rk{2i}}{\Rk{\kappa+1}}[j].
\end{equation}

Since any individual term in (\ref{Eq. 2M-MHCompeleteSignalEven}), $\Nob{1}{\Rk{2i}}{\Rk{\kappa+1}}[j]$, is a Poisson random variable with time-varying mean, $\NobTwo{1}{\Rk{\kappa+1}}[j]$ is also a Poisson random variable whose mean is the sum of the means of all individual variables, and can be written as 
\begin{equation}
	\label{Eq. 2M-MHCompeleteSignalEvenMean} 
	\NobETwo{1}{\Rk{\kappa+1}}[j] = \sum_{i = 0}^{\lfloor \frac{Q+1}{2} \rfloor} \NobE{1}				{\Rk{2i}}{\Rk{\kappa+1}}[j].
\end{equation}

Analogously, when $\kappa +1$ is even, the complete received signal at $\Rk{\kappa+1}$ in the $j$th bit interval is 
\begin{equation}
	\label{Eq. 2M-MHCompeleteSignalOdd} 
	\NobTwo{2}{\Rk{\kappa+1}}[j] = \sum_{i = 0}^{\lfloor \frac{Q-1}{2} \rfloor} \Nob{2}{\Rk{2i+1}}{\Rk{\kappa+1}}[j],
\end{equation}
which is a Poisson random variable whose time-varying mean is the sum of the means of all individual variables, i.e., 
\begin{equation}
	\label{Eq. 2M-MHCompeleteSignalOddMean}
	\NobETwo{2}{\Rk{\kappa+1}}[j] = \sum_{i = 0}^{\lfloor \frac{Q-1}{2} \rfloor} \NobE{2}{\Rk{2i +1}}{\Rk{\kappa+1}}[j].
\end{equation}

Finally, the expected error probability of the considered network can be evaluated via (\ref{Eq. ErrorExpSeqmulti-hop}), after substituting $\NobTwo{\kappa+1}{\Rk{\kappa+1}}[j]$ and $\xi_{\Rk{\kappa+1}}$ with $\NobTwo{}{\Rk{\kappa+1}}[j]$ and $\xi_{\Rk{\kappa+1}}^{BI}[j]$, respectively, and considering that all conditional probabilities have to be conditioned on the set $\{ \wsh{\Rk{}}{1}{j+ \kappa -2} \}, \forall \Rk{} \in \mathcal{R_E}$, when $\kappa+1$ is odd, and on $\{ \wsh{\Rk{}}{1}{j+ \kappa -2} \}, \forall \Rk{} \in \mathcal{R_O}$, when $\kappa+1$ is even. The detected bits in $\wsh{\Rk{\kappa+1}}{1}{j+\kappa -2}$, i.e., $\wh{\Rk{\kappa+1}}{i}, i < j + \kappa -2$, are modeled as $\wh{\Rk{\kappa+1}}{i} = |\lambda - \wt{S}{i}|$, where $\lambda \in \{ 0,1 \}$ is the outcome of a coin toss with $\pr(\lambda=1)= P_{e_{\kappa}}[i|\{ \wsh{\Rk{}}{1}{i+ \kappa -2} \}] $ and $\pr(\lambda =0)=1-\pr(\lambda=1)$.             
\section{SINGLE-MOLECULE MULTI-HOP NETWORK} 
\label{Sec. SM-MH} 
We now consider an SM-MH network where the same type of molecule is employed in all hops, cf. Fig. \ref{Fig.SysMod}. In this scenario, in addition to the occurrence of backward-ISI and forward-ISI, utilizing the same type of molecule for transmission and reception at relay $\Rk{\kappa}$ leads to the occurrence of self-interference. In particular, some of the molecules released by relay node $\Rk{\kappa}$ at the beginning of a bit interval stay nearby and are observed during this bit interval and in subsequent bit intervals inside $V_{\Rk{\kappa}}$. This effect causes \textit{self-interference}. In the following, we first study the effect of self-interference for a two-hop network, where the effect of backward-ISI does not exist, and we propose two approaches to mitigate the self-interference. Then, we consider a multi-hop network and extend our proposed algorithms to the joint mitigation of self-interference and backward-ISI. 
\subsection{Two-hop Network} In this case, node $S$ releases molecules of type $A_{1}$, which are detected by relay node $\Rk{1}$. Node $\Rk{1}$ also releases molecules of type $A_1$ to forward the detected message to node $D$. We first consider full-duplex transmission. We provide a short example to clarify the occurrence of self-interference. 

\textit{Example 2:} Let us assume that the information sequence emitted by node $S$ is ``10'', i.e., $\wt{S}{j}=1, \wt{S}{j+1}=0, \text{and } \wt{S}{m}=0, m<j$, and that no error occurs in the transmission of $\wt{S}{j}$ to node $\Rk{1}$, i.e., $\wh{\Rk{1}}{j}=\wt{S}{j}=1$. At the beginning of the $(j+1)$th bit interval, node $\Rk{1}$ releases $N_{A_{1}}$ molecules to forward the detected message to node $D$, i.e., $\wt{\Rk{1}}{j+1}=1$. Due to the random movement of the molecules, some of the molecules released by the relay node may be observed within its own volume, $V_{\Rk{1}}$, at the time of sampling for detection of $\wt{S}{j+1}$. This self-interference may lead to an erroneous decision for $\wh{\Rk{1}}{j+1}$.

In the following, we propose two approaches to mitigate self-interference: 1) employing an adaptive decision threshold at the relay, and 2) employing half-duplex relaying instead of full-duplex relaying. 

\subsubsection{Adaptive Decision Threshold} In the first approach, analogous to the scheme for the mitigation of backward-ISI described in Section \ref{SubSec.Backward-ISI}, the relay adjusts its decision threshold in each bit interval based on the information bits it has detected in all previous bit intervals, $\wsh{\Rk{1}}{1}{j-1}$. Thus, the adaptive decision threshold of the relay in the $j$th bit interval, $\xi_{\Rk{1}}^{SI}[j]$, can be written as 
\begin{equation}
	\label{Eq. Adp Thresh1} 
		\xi_{\Rk{1}}^{SI}[j]=\xi + \xi_{Exp}^{SI}[j], 
\end{equation}
where $\xi_{Exp}^{SI}[j]$ changes adaptively based on the number of molecules \emph{expected} within $V_{\Rk{1}}$, given $\wsh{\Rk{1}}{1}{j-1}$. To optimize $\xi_{Exp}^{SI}[j]$, we have to determine the probability of observing a given molecule transmitted by the relay node $\Rk{1}$ at $t=0$ within $V_{\Rk{1}}$ at time $t$. We denote this probability as $\Pob{1}{\Rk{1}}{\Rk{1}}(t)$. $\Pob{1}{\Rk{1}}{\Rk{1}}(t)$ may be considered as a special case of $\Pob{}{\xstar}{\ystar}(t)$ when $\xstar = \ystar$, i.e., $\vec{r}_{\ystar} = \vec{r}_{\xstar}$. However, in this case, the conditions necessary for the validity of the uniform concentration assumption do not hold \cite{NoelPro1}. Hence, we can not use (\ref{Eq. Pob}) to evaluate $\Pob{1}{\Rk{1}}{\Rk{1}}(t)$. The general form of $\Pob{}{\xstar}{\ystar}(t)$, when the uniform concentration assumption is not made, is given by \cite[Eq. (27)]{NoelPro1}. It can be shown that, by using l'H$\hat{\text{o}}$pital's rule, $\Pob{1}{\Rk{1}}{\Rk{1}}(t)$ in the limit of  $|\vec{r}_{\ystar} - \vec{r}_{\xstar}| \rightarrow 0$ can be written as 
\begin{equation}
	\label{Eq. Pselfobs} 
	\Pob{1}{\Rk{1}}{\Rk{1}}(t) = \text{erf}\left(  \frac{r_{\Rk{1}}}{2 \hspace{1mm}\sqrt[]{D_{A_{1}}t}} \right) - \frac{ r_{\Rk{1}} \exp{\left( \frac{{-r_{\Rk{1}}}^{2}}{4D_{A_{1}}t} \right)}}{\sqrt[]{\pi D_{A_{1}}t}}, 
\end{equation}
where $r_{\Rk{1}}$ is the radius of the relay node $\Rk{1}$, and $\text{erf}(\cdot)$ denotes the error function as defined by \cite[Eq. (7.1.1)]{abramowitz}. Thus, given $\wsh{\Rk{1}}{1}{j-1}$, the expected number of molecules observed within $V_{\Rk{1}}$ in the $j$th bit interval, $\NobE{1}{\Rk{1}}{\Rk{1}}[j]$, can be written as 
\begin{equation} 
	\label{Eq. ExpSelfOb} 
	\NobE{1}{\Rk{1}}{\Rk{1}}[j] = N_{A_{1}} \sum_{i=1}^{j-1} \wh{\Rk{1}}{i} \sum_{m=1}^{M} \Pob{1}{\Rk{1}}{\Rk{1}}((j-i)T+t_{m}), 
\end{equation} 
and the varying part of the adaptive decision threshold of the relay becomes $\xi_{Exp}^{SI}[j] = \NobE{1}{\Rk{1}}{\Rk{1}}[j]$. The number of molecules observed inside $V_{\Rk{1}}$ in the $j$th bit interval when only the relay node transmits, $\Nob{1}{\Rk{1}}{\Rk{1}}[j]$, is a Poisson random variable with the mean given by (\ref{Eq. ExpSelfOb}). The complete received signal at the relay node in the $j$th bit interval, $\NobTwo{1}{\Rk{1}}[j]$, is the sum of two signals, i.e., $\NobTwo{1}{\Rk{1}}[j] = \Nob{1}{S}{\Rk{1}}[j] + \Nob{1}{\Rk{1}}{\Rk{1}}[j]$. Since $\Nob{1}{S}{\Rk{1}}[j]$ and $\Nob{1}{\Rk{1}}{\Rk{1}}[j]$ are Poisson random variables with time-varying means, $\NobTwo{1}{\Rk{1}}[j]$ is also a Poisson random variable whose mean is the sum of the means of the individual variables, i.e., $\NobETwo{1}{\Rk{1}}[j] = \NobE{1}{S}{\Rk{1}}[j] + \NobE{1}{\Rk{1}}{\Rk{1}}[j] $. Analogously, the complete received signal at node $D$ in the $j$th bit interval, $\NobTwo{1}{D}[j]$, is the sum of two Poisson random variables $\Nob{1}{S}{D}[j]$ and $\Nob{1}{\Rk{1}}{D}[j]$. Thus, $\NobTwo{1}{D}[j]$ is also a Poisson random variable with time-varying mean $\NobETwo{1}{D}[j] = \NobE{1}{S}{D}[j] + \NobE{1}{\Rk{1}}{D}[j]$. Finally, the expected error probability of the considered network can be evaluated via (\ref{Eq. ErrorExpSeq}), after substituting $\xi_{R}$ with $\xi_{\Rk{1}}^{SI}[j]$, and considering that all conditional probabilities in (\ref{Eq. ErrorExpSeq}) have to be conditioned on both $\ws{S}{1}{j-1}$ and $\wsh{\Rk{1}}{1}{j-1}$. The detected bits in $\wsh{\Rk{1}}{1}{j-1}$, i.e., $\wh{\Rk{1}}{i}, i<j$, are modelled as $\wh{\Rk{1}}{i} = |\lambda-\wt{S}{i}|$, where $\lambda \in \{0,1\}$ is the outcome of a coin toss with $\pr(\lambda=1)= P_{e_{1}}[i|\ws{S}{1}{i-1},\wsh{\Rk{1}}{1}{i-1}]$ and $\pr(\lambda=0) = 1 - \pr(\lambda =1)$. $P_{e_{1}}[i|\ws{S}{1}{i-1},\wsh{\Rk{1}}{1}{i-1}]$ can be evaluated via (\ref{Eq. ExpErrorOneHop}), after substituting $\Nob{}{\xstar}{\ystar}[i]$ and $\xi_{\ystar}$ with $\NobTwo{1}{\Rk{1}}[i]$ and $\xi_{\Rk{1}}^{SI}[i]$, respectively. 

\subsubsection{Half-Duplex Relaying} In the second approach to mitigate self-interference, half-duplex relaying is adopted. In half-duplex relaying, reception and transmission at the relay occur in two consecutive bit intervals, giving the molecules released at the relay node time to leave $V_{\Rk{1}}$, such that they are less likely to interfere with the relay's decisions. 

For half-duplex relaying, the nodes communicate as follows. In odd bit intervals, $S$ transmits and $\Rk{1}$ receives, and in even bit intervals, $\Rk{1}$ transmits and $D$ receives. In other words, in the $(2j-1)$th bit interval, node $S$ transmits the $j$th information bit, i.e., $\wt{S}{2j-1}$, which is detected by node $\Rk{1}$ as $\wh{\Rk{1}}{2j-1}$, and in the $(2j)$th bit interval, node $\Rk{1}$ transmits the $j$th bit detected in the previous bit interval, i.e., $\wt{\Rk{1}}{2j} = \wh{\Rk{1}}{2j-1}$. This bit is then detected at node $D$ as $\wh{D}{2j}$. 

The expected error probability for half-duplex relaying can be evaluated via (\ref{Eq. ErrorExpSeq}), after substituting $\wt{S}{j}$ and $\wt{R}{j+1}$ with $\wt{S}{2j-1}$ and $\wt{R}{2j}$, respectively, and considering that all conditional probabilities have to be conditioned on both $\ws{S}{1}{2j-1}$ and $\wsh{\Rk{1}}{1}{2j-1}$, where $\wt{S}{2i} = \wh{\Rk{1}}{2i}= 0 $ for $i \in \{1,2,...,(j-1) \}$. 

\subsection{Multi-hop Network} In a multi-hop network using the same type of molecule in all hops, the complete received signal at node $\Rk{\kappa}$ in the $j$th bit interval can be written as 
\begin{equation}
	\label{Eq. ComSigMul_Sametype} 
	\NobTwo{1}{\Rk{\kappa}}[j] = \sum_{\omega=0}^{Q} \Nob{1}{\Rk{\omega}}{\Rk{\kappa}}[j],
\end{equation} 
which is a Poisson random variable whose time-varying mean is the sum of the means of all individual variables, i.e., $\NobETwo{1}{\Rk{\kappa}}[j] = \sum_{\omega=0}^{Q} \NobE{1}{\Rk{\omega}}{\Rk{\kappa}}[j]$. In order to jointly mitigate the effects of self-interference and backward-ISI, we combine the two proposed schemes to mitigate the self-interference with the scheme proposed in Section \ref{Sec. TM-MH}  to mitigate the backward-ISI as follows. For full-duplex transmission, given $\wsh{\Rk{\kappa}}{1}{j-1}$, the relay node $\Rk{\kappa}$ adjusts its decision threshold in the $j$th bit interval as 
\begin{equation}
	\label{Eq. AdpativeThresholdBISIFD} 
	\xi_{\Rk{\kappa},FD}^{SI,BI}[j] = \xi + \xi_{Exp}^{SI}[j] + \xi_{Exp}^{BI}[j], 
\end{equation}
where $\xi_{Exp}^{BI}[j]$ can be obtained from (\ref{Eq. ExpBIOb}), and $\xi_{Exp}^{SI}[j]$ can be obtained from (\ref{Eq. ExpSelfOb}) after substituting $\Rk{1}$ with $\Rk{\kappa}$. 

For half-duplex transmission, the relay node $\Rk{\kappa}$ adjusts its decision threshold as 
\begin{equation}
	\label{Eq. AdaptveThresholdBISIHD}  
	\xi_{\Rk{\kappa},HD}^{BI}[l] = \xi + \xi_{Exp}^{BI}[l],	
\end{equation} 
where $l=2j$ if the index of the rely node is even, and $l=2j-1$ if the index of the relay node is odd. $\xi_{Exp}^{BI}[l]$ can be evaluated via (\ref{Eq. ExpBIOb}), given $\wsh{\Rk{\kappa}}{1}{l-1}$. 

The expected error probability of both of the above-mentioned protocols can be evaluated via (\ref{Eq. ErrorExpSeqmulti-hop}), after substituting $\NobTwo{\kappa+1}{\Rk{\kappa+1}}[j+\kappa]$ with the complete received signal given in (\ref{Eq. ComSigMul_Sametype}), i.e., $\NobTwo{1}{\Rk{\kappa+1}}[j+\kappa]$ for full-duplex transmission ($\NobTwo{1}{\Rk{\kappa+1}}[l+\kappa]$ for half-duplex transmission), $\xi_{\Rk{\kappa+1}}$ with $\xi_{\Rk{\kappa+1},FD}^{SI,BI}[j+\kappa]$ for full-duplex transmission ($\xi_{\Rk{\kappa+1},HD}^{BI}[l+\kappa]$ for half-duplex transmission), and considering that all conditional probabilities have to be conditioned on $\{ \wsh{\Rk{\omega}}{1}{j+ \kappa -2} \}, \omega \in \{0,...,Q \}$. The previously detected bits in $\wsh{\Rk{\kappa}}{1}{j+\kappa -2}$, i.e., $\wh{\Rk{\kappa}}{i}, i \leq j + \kappa -2$, are modeled as $\wh{\Rk{\kappa}}{i} = |\lambda - \wt{S}{i}|$, where $\lambda \in \{ 0,1 \}$ is the outcome of a coin toss with $\pr(\lambda=1)= P_{e_{\kappa}}[i|\{ \wsh{\Rk{\omega}}{1}{i+ \kappa -2} \}] $ and $\pr(\lambda =0)=1-\pr(\lambda=1)$ for full-duplex transmission. For half-duplex transmission, we model the detected bits as $\wh{\Rk{\kappa}}{l^{'}} = |\lambda - \wt{S}{2i}|, i< \lfloor \frac{l+\kappa-2}{2} \rfloor$, where $l^{'}=2i$ if $\kappa$ is even, and $l^{'}=2i-1$ if $\kappa$ is odd, and 
$\pr(\lambda=1) = P_{e_{\kappa}}[i|\{ \wsh{\Rk{\omega}}{1}{2i+ \kappa -1} \}$ and $\pr(\lambda =0)=1-\pr(\lambda=1)$. 
\section{NUMERICAL RESULTS}
\label{Sec.NumRes} 
In this section, we present simulation and analytical results to evaluate the performance of the proposed relaying schemes. We also show the excellent match between our simulation and analytical results. We adopted the particle-based stochastic simulator introduced in \cite{NoelJ1}. In our simulations, time is advanced in discrete steps $t_{0}$, i.e., the time between two consecutive samples, where in each time step molecules undergo random movement. The environment parameters are listed in Table \ref{Table2}.
 
In order to focus on the comparison of the performance of the different relaying protocols, we keep the physical parameters of the relays and the destination node constant throughout this section. In particular, we assume that $ r_{\Rk{\kappa}} = r_{D}$, and nodes $\Rk{\kappa}$, $ 1 \leq \kappa \leq Q$, and $S$ release the same numbers of molecules. The only parameters that we vary are the numbers of relays, the decision threshold, the modulation bit interval, the frequency of sampling, and the number of released molecules. We assume $x_{D} = 1$ $\mu m$ throughout this section, unless specified otherwise.

\begin{table}
\renewcommand{\arraystretch}{1}
\caption{SYSTEM PARAMETERS USED IN SIMULATIONS}
\label{Table2}
\centering
\begin{tabular}{|c|c|c|} 
\hline 
\bfseries Parameter & \bfseries Symbol  & \bfseries Value \\ 
\hline 
Probability of binary 1 & $P_1$ & 0.5 \\ 
\hline 
Length of transmitter sequence & $L$ & 50 \\ 
\hline 
Radius of rely node $\Rk{\kappa}$ & $r_{\Rk{\kappa}}$ & 45 nm \\ 
\hline 
Radius of node $D$ & $r_{D}$ & 45 nm \\ 
\hline 
Diffusion coefficient \cite{NoelJ1},\cite{NoelJ2} & $D_{A_{f}}$ & $4.365 \times 10^{-10} \frac{\text{m}^2}{\text{s}}$ \\
\hline
\end{tabular} 
\end{table}
\begin{table}
\renewcommand{\arraystretch}{1}
\caption{Summary of the Considered Relaying Protocols}
\label{Table3}
\centering
\begin{tabular}{|c|c|c|c|} 
\hline 
\bfseries Relaying &\bfseries Relaying &  \bfseries Relay & \bfseries Protocol \\
\bfseries Scheme &\bfseries Mode &  \bfseries Detection & \bfseries Acronym \\ 
& & \bfseries Threshold & \\
\hline 
MM-MH & Full-duplex &  $\xi$ & FD \\ 
\hline 
2M-MH & Full-duplex &  $\xi$ & FD \\ 
\hline 
2M-MH & Full-duplex &  (\ref{Eq.AdpThreshBI}) & FD-A \\ 
\hline 
SM-MH & Full-duplex &  $\xi$ & FD \\ 
\hline
SM-MH & Full-duplex &  (\ref{Eq. Adp Thresh1}) & FD-A-SI \\
\hline
SM-MH & Half-duplex &  $\xi$ & HD \\
\hline 
SM-MH & Full-duplex &  (\ref{Eq. AdpativeThresholdBISIFD}) & FD-A-BI-SI \\ 
\hline 
SM-MH & Half-duplex &  (\ref{Eq. AdaptveThresholdBISIHD}) & HD-A-BI \\ 
\hline 
\end{tabular}
\end{table} 
In Table \ref{Table3}, we summarize the protocols considered for multi-hop transmission. The FD protocol, i.e., the full-duplex mode without an adaptive threshold in 2M-MH and SM-MH, is mainly considered for comparison to illustrate the effects of backward-ISI and self-interference, respectively. 

In the following, we refer to the case when no relay is deployed between node $S$ and node $D$ as the \textit{baseline case}. For a fair comparison between the multi-hop case and the baseline case, we assume that for multi-hop transmission each relay node $\Rk{\kappa}, \kappa \in \{1,...,Q \}$, and node $S$ release $N_{A_1}/(Q+1)$ molecules, respectively, to transmit information bit ``1'', where $N_{A_1}$ is the number of molecules released by node $S$ in the baseline case. In all figures, the expected error probability of the multi-hop link was evaluated via (\ref{Eq. ErrorExpSeqmulti-hop}), after taking into account the modifications required for each protocol.

\subsection{Single-Hop Optimization} 

\begin{figure}[t]
	\centering\vspace*{-6mm}
	\includegraphics[width=3.7 in]{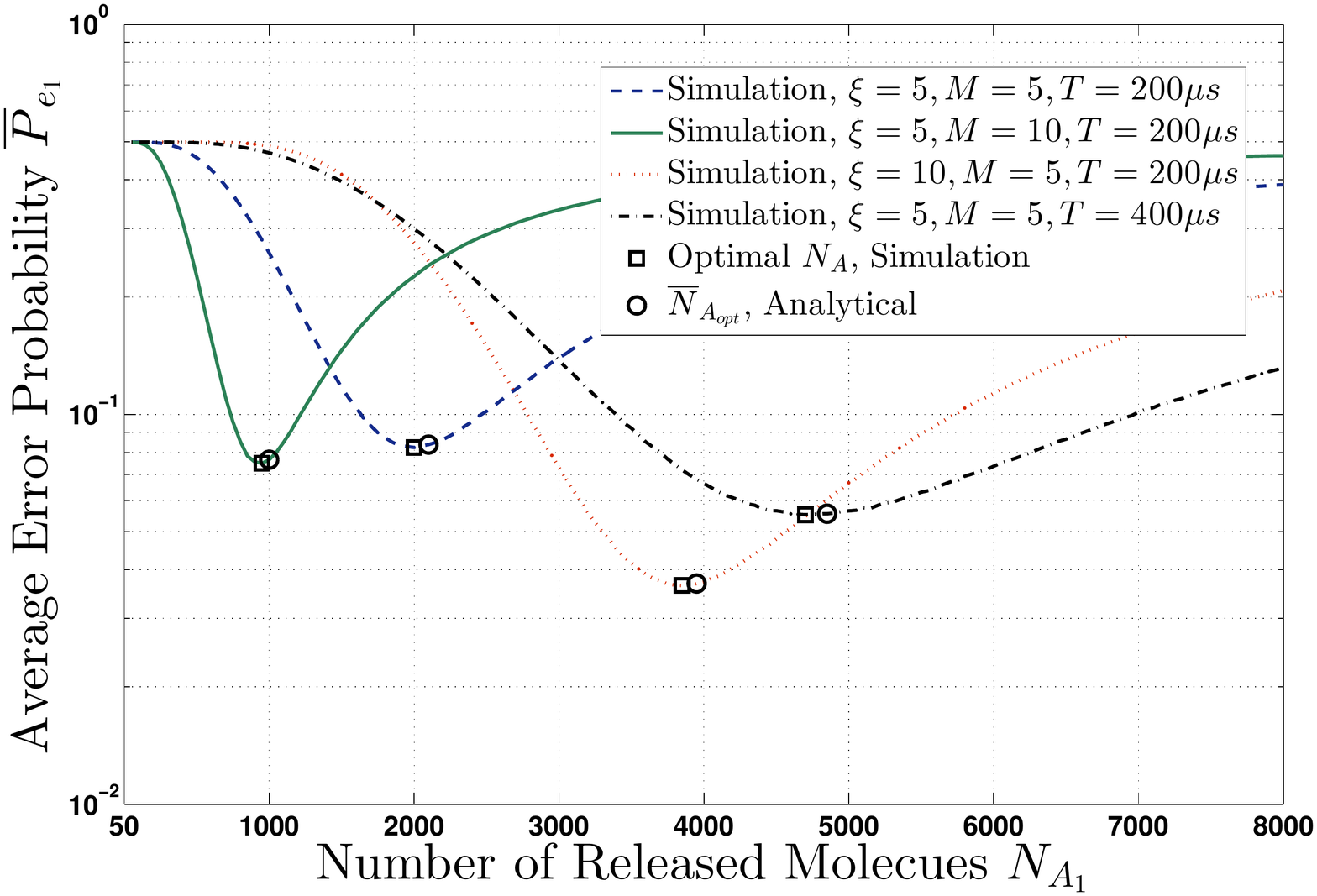}\vspace*{-6mm}
	\caption{Average error probability of a single link as a function of the number of released molecules $N_{A_1}$.} \label{Fig.11} 
\end{figure} 
\begin{figure}
	\centering\vspace*{-6mm}
	\includegraphics[width=3.7 in]{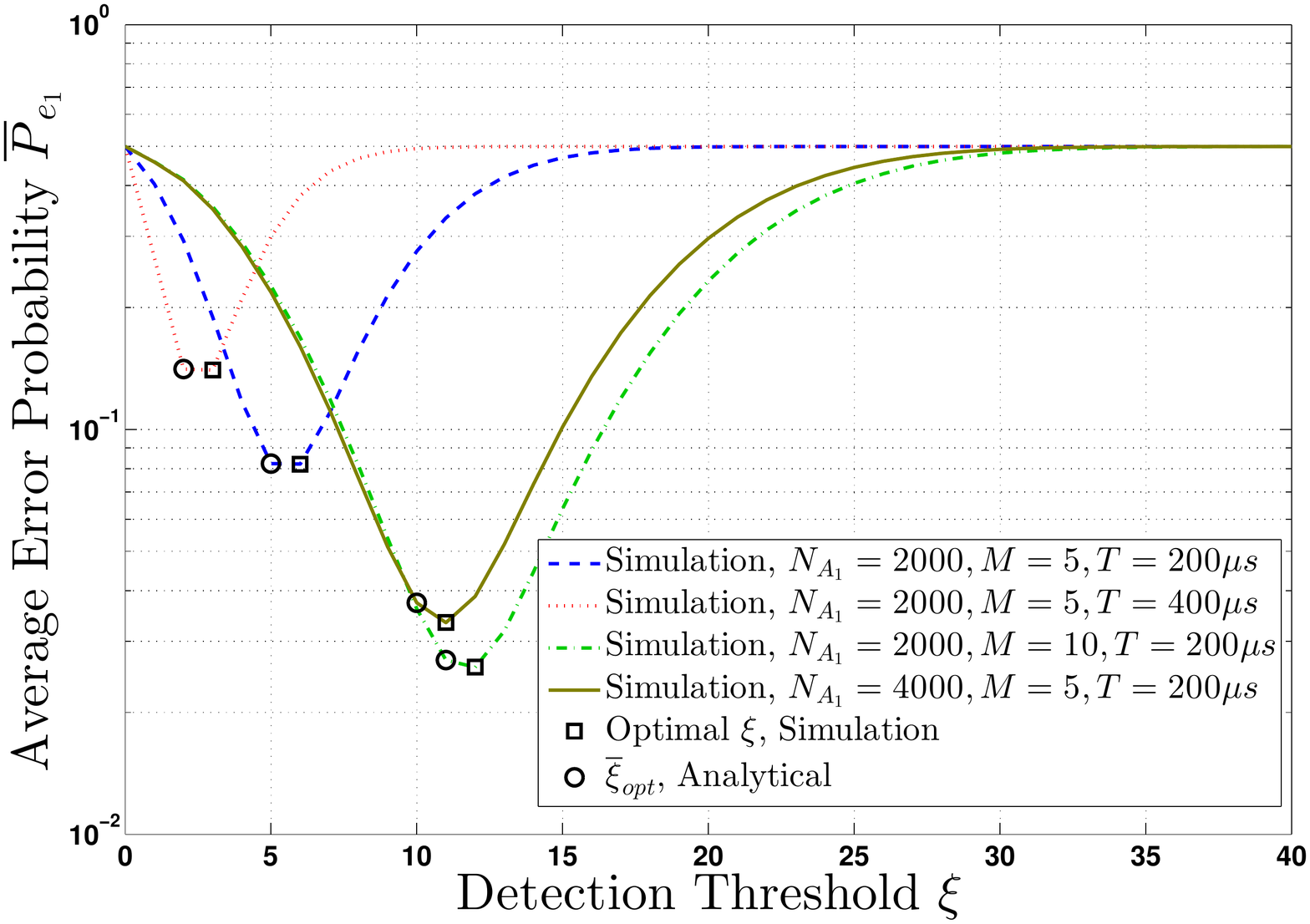}\vspace*{-6mm}
	\caption{Average error probability of a single link as a function of the detection threshold $\xi$.} \label{Fig.12}
\end{figure} 
In Figs. \ref{Fig.11} and \ref{Fig.12}, we evaluate the expected error probability of a \emph{single} link as a function of $N_{A_{1}}$ and $\xi$, respectively, to assess the accuracy of the optimal values, $\overline{N}_{A_{opt}}$ and $\overline{\xi}_{opt}$, which are derived in Section \ref{Sec.Two-hopNetPerAna}. We assume that $x_{D} = 250$ nm and evaluate $\overline{P}_{e_{1}}$ for different system parameters, i.e., $M = \{5,10\}$, $T = \{200, 400\}$ $\mu$s, and $\xi = \{5,10\}$ in Fig. \ref{Fig.11}, and $N_{A_{1}} = \{2000,4000\}$ in Fig. \ref{Fig.12}. We emphasize that the results shown in Figs. \ref{Fig.11} and \ref{Fig.12} are also valid for the expected error probability of individual hops of an MM-MH network (when the relative distance between two adjacent nodes is 250 nm). 

In Fig. \ref{Fig.11}, we observe that, by doubling $\xi$, the optimal $N_{A_{1}}$ is approximately doubled which is in agreement with (\ref{Eq. Optimum Na p1=p0}). We can also see that increasing $M$ and $T$ decreases and increases the optimal $N_{A_{1}}$, respectively. Fig. \ref{Fig.12} shows that, for a given $N_{A_{1}}$, by increasing the number of samples $M$ per bit interval the optimal $\xi$ increases. Furthermore, we observe that by increasing the modulation bit interval $T$, due to the decreasing ISI, the optimal $\xi$ also decreases. Finally, we note the excellent match between simulation and analytical results.       

\subsection{Multi-Molecule Multi-Hop Network}
\begin{figure}[t]
	\centering\vspace*{-6mm}
	\includegraphics[width=3.7 in]{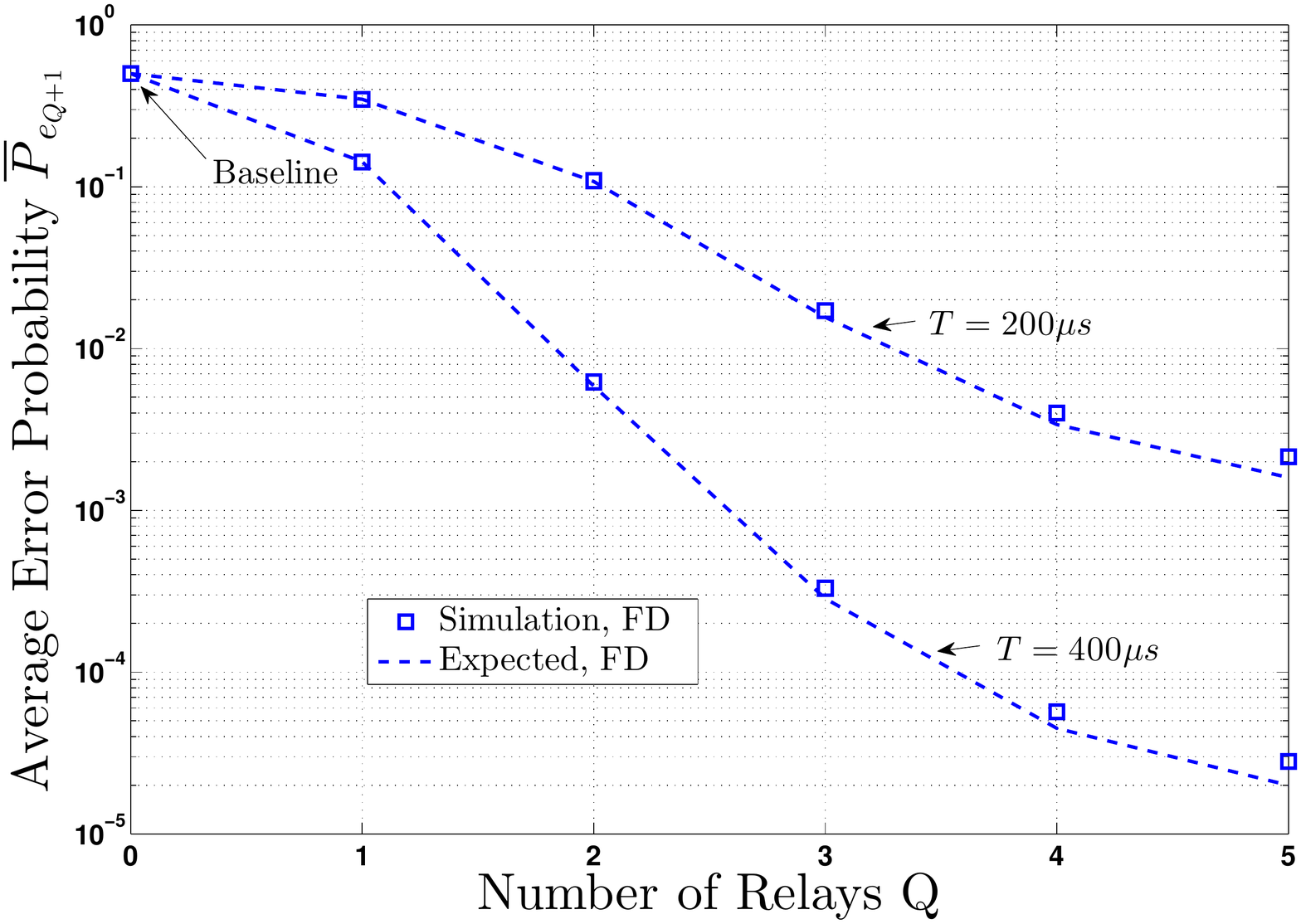}\vspace*{-6mm}
	\caption{Average error probability of an MM-MH network as a function of the number of relays. $Q=0$ is the baseline case.}
	\label{Fig.2} 
\end{figure}
In Fig. \ref{Fig.2}, we evaluate the performance of MM-MH networks as a function of the number of relays deployed between node $S$ and node $D$. We use a different type of molecule in each hop and employ the FD protocol. Node $D$ is placed at $x_{D} = 1 \text{ } \mu \text{m}$, and $N_{A_{1}} = 20000$ for the baseline case. We set the number of samples per bit interval to $M = 10$ with $t_{0} = 20 \text{ } \mu \text{s}$. For a fair comparison of the performance of the network for different bit intervals, we assume that the frequency of sampling, $t_{0}$, and the number of samples per bit interval are independent of $T$, i.e., for any $T$ the samples are taken at times $t = \{ 20, 40, 60, \ldots, 200 \}$ $\mu$s within the current bit interval. Furthermore, for the multi-hop case and the baseline case, we chose the average optimal detection threshold, $\overline{\xi}_{opt}$, given $T$, $M$, the relative distance between two adjacent relay nodes, and the number of molecules released by each relay, such that the average error probability of the individual hops is minimized. The results in Fig. \ref{Fig.2} show that by increasing the number of relay nodes between node $S$ and node $D$, the overall performance of the network improves. This is because, by increasing the number of relay nodes, the relative distance between two adjacent relay nodes decreases which leads to an improvement in the performance of individual hops. We also observe that by increasing the bit interval $T$, the performance improves by orders of magnitude, especially for a large number of relays. This shows that if the number of the types of molecules that can be used is not limited, ISI is the dominant performance limiting factor of the network. We also note the excellent agreement between simulation and analytical results which confirms the accuracy of the approximations made for evaluation of the expected error probability of the multi-hop network. 

\subsection{Two-Molecule Multi-hop Network}
\begin{figure}[t]
 \centering\vspace*{-6mm}
\includegraphics[width=3.7 in]{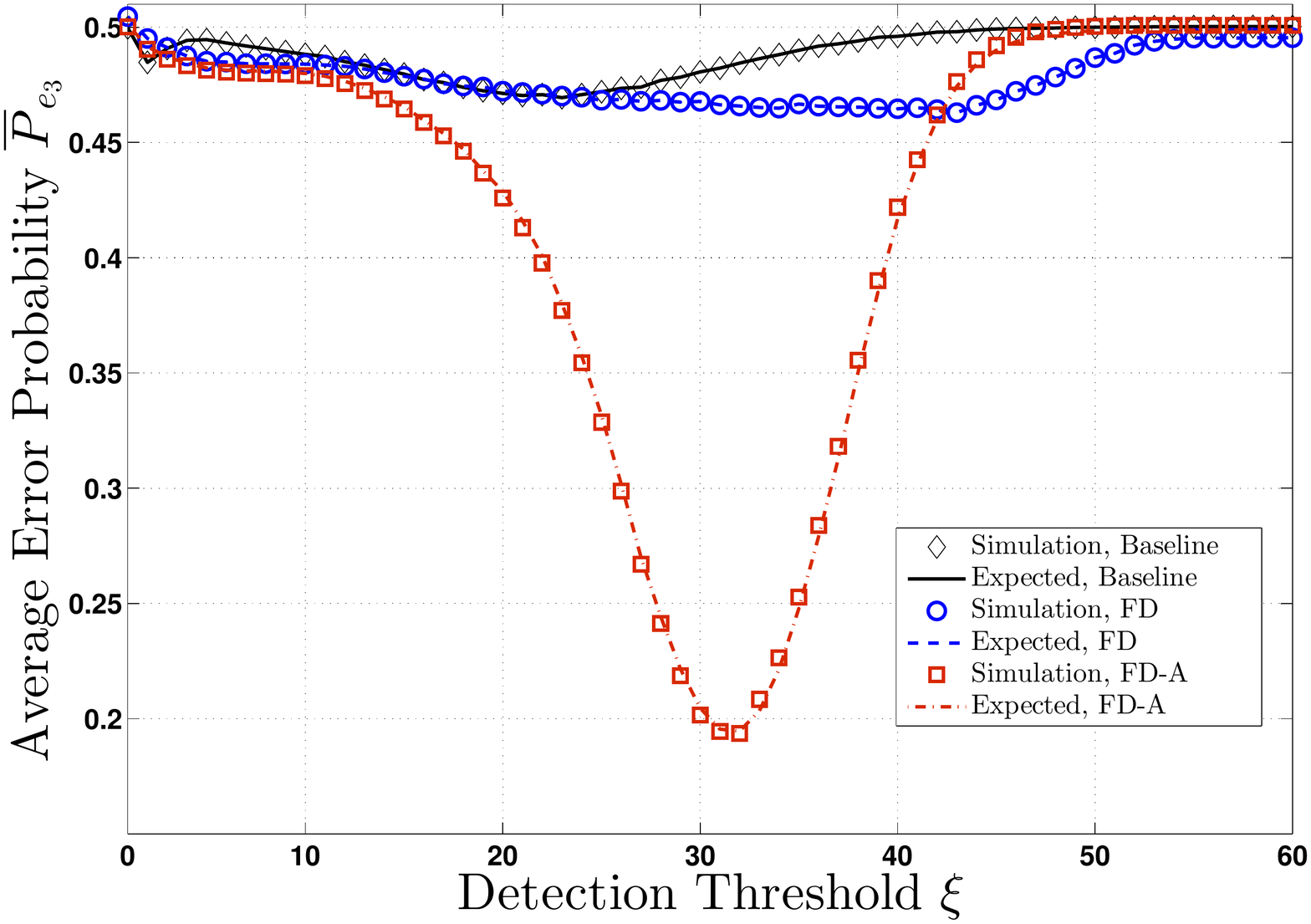}\vspace*{-6mm}
\caption{Average error probability of a 2M-MH network as a function of the detection threshold for $Q=2$.} \label{Fig.3}
\end{figure}

\begin{figure}
\centering\vspace*{-6mm}
\includegraphics[width=3.7 in]{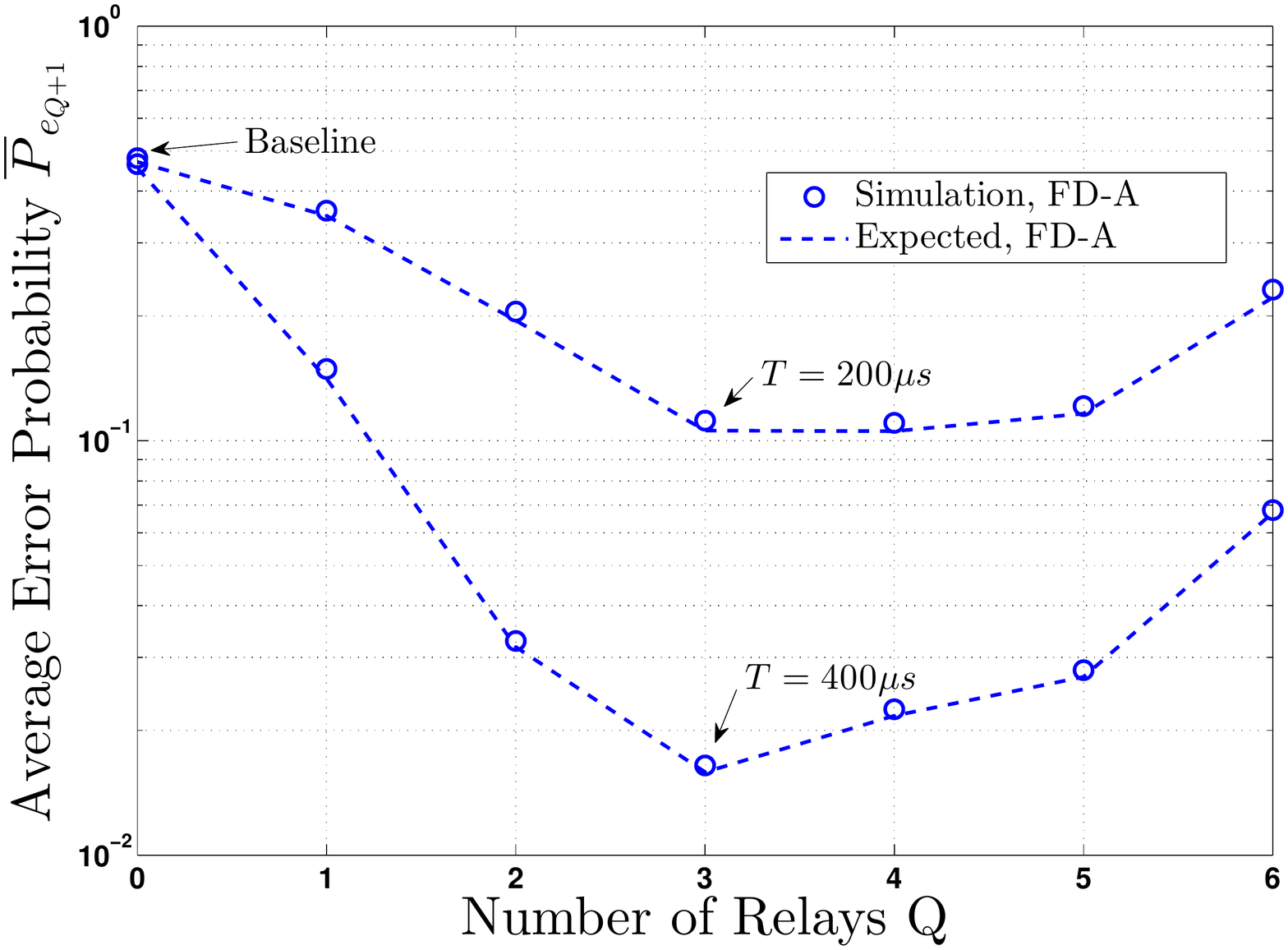}\vspace*{-6mm}
\caption{Average error probability of a 2M-MH network as a function of the number of relays. $Q=0$ is the baseline case.} \label{Fig.4}
\end{figure}

We now consider a 2M-MH network and study the effects of backward-ISI and forward-ISI. In Figs \ref{Fig.3} and \ref{Fig.4}, we adopt $x_D= 1$ $\mu$m, $M=10$, and $t_0 = 20$ $\mu$s. 
 
Fig. \ref{Fig.3} shows the performance of a three-hop network as a function of the detection threshold for $T = 200$ $\mu$s. This is the smallest multi-hop network where backward-ISI occurs at the first relay node. We compare the performance of the baseline case with the FD and FD-A protocols. For FD and the baseline case, we adopted $\xi_D = \xi_{\Rk{1}} = \xi_{\Rk{2}} = \xi$, and for the FD-A protocol, the fixed part of the adaptive threshold at $\Rk{1}$ is $\xi$. The simulation results reveal that the occurrence of backward-ISI at the first relay heavily affects the performance of the first relay in the FD protocol, and as a result, the overall performance of the network is limited by the performance of the first hop. In fact, the FD protocol has almost the same performance as the baseline case, where no relay is used. However, the proposed FD-A protocol can effectively mitigate the backward-ISI and performs significantly better than the baseline case.
  
In Fig. \ref{Fig.4}, we investigate the performance of a 2M-MH network as a function of the number of relays deployed between node $S$ and node $D$. We compare the performance of the baseline case ($Q=0$) with the FD-A protocol. We numerically found the optimal detection threshold that minimizes the expected error probability of the overall network. The results in Fig. \ref{Fig.4} show that, by deploying more relays between node $S$ and node $D$, the performance of the 2M-MH network improves first and then deteriorates. This is because when the number of deployed relays between node $S$ and node $D$ increases, the distance between adjacent relays decreases. This is beneficial at first but also causes increased levels of forward-ISI which eventually becomes the performance limiting factor of the network.

\subsection{Single-Molecule Multi-hop Network} 
\begin{figure}[t]
 \centering\vspace*{-6mm}
\includegraphics[width=3.7 in]{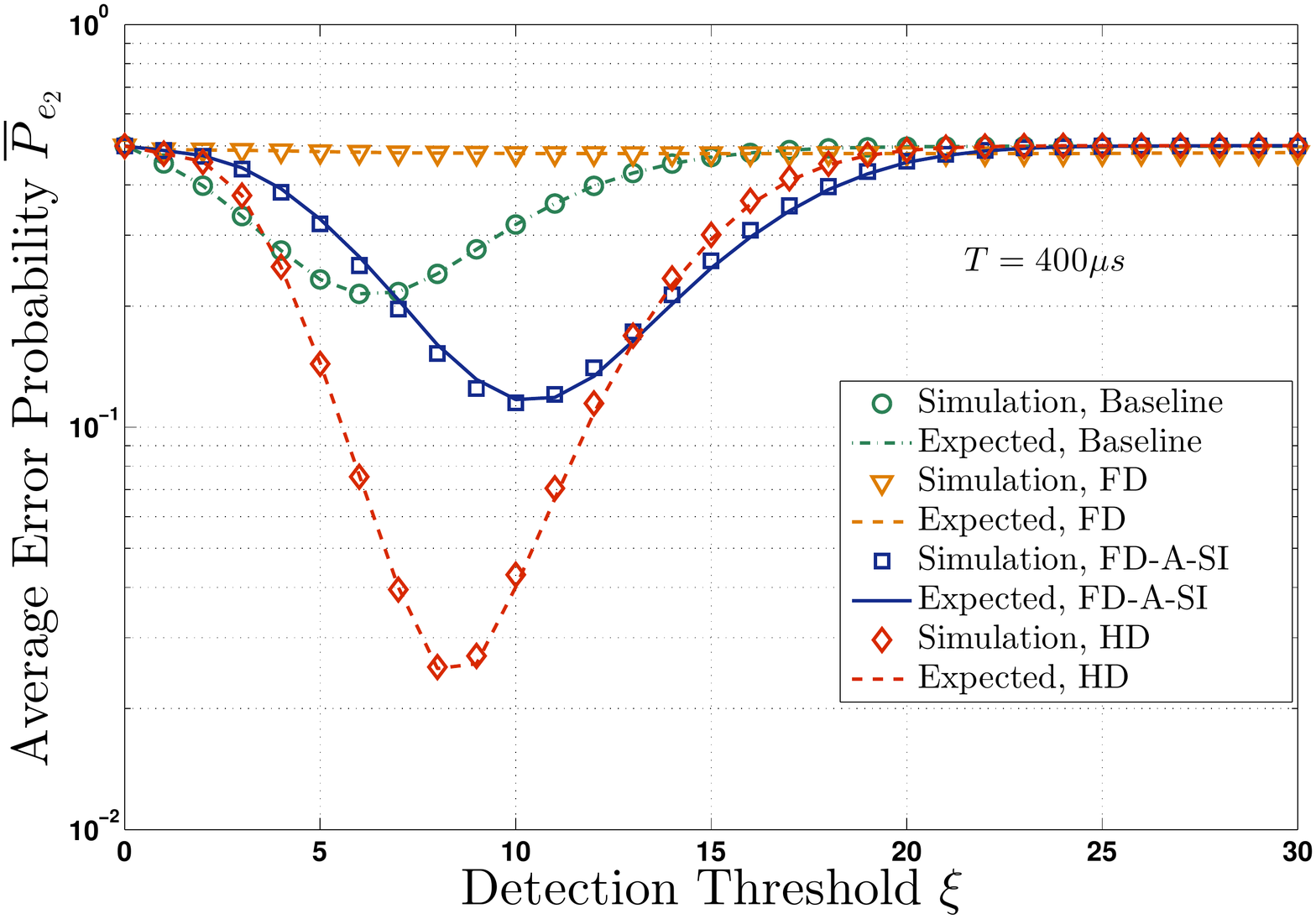}\vspace*{-6mm}
\caption{Average error probability of a SM-MH network as a function of the detection threshold, where $Q=1$.} \label{Fig.5}
\end{figure}

Fig. \ref{Fig.5} shows the average error probability of SM-MH as a function of the detection threshold for $Q=1$ and $T = 400$ $\mu$s. We show the performance of FD, HD, FD-A-SI, and the baseline case for $x_D= 600$ nm, $M=5$, $t_0 = 20$ $\mu$s, and $N_{A_{1}}=10000$ (for the baseline case). For FD, HD, and the baseline case, we adopt $\xiN{D}=\xiN{\Rk{1}}=\xi$, and for the FD-A-SI protocol, the fixed part of the adaptive threshold is equal to $\xi$. Fig. \ref{Fig.5} reveals that the FD protocol performs even worse than the baseline case. This confirms the performance-limiting effect of self-interference. However, the proposed FD-A-SI and HD protocols are effective in mitigating self-interference and perform better than the baseline scheme. Furthermore, the HD protocol performs better than the FD-A protocol. This is because, for the FD-A protocol, the decision threshold can only be adapted based on the expected number of observed molecules, which may differ from the actual number of observed molecules. We note that the better performance of the HD protocol comes at the expense of decreasing the transmission rate by a factor of two.

\begin{figure}[t]
\centering\vspace*{-6mm}
\includegraphics[width=3.7 in]{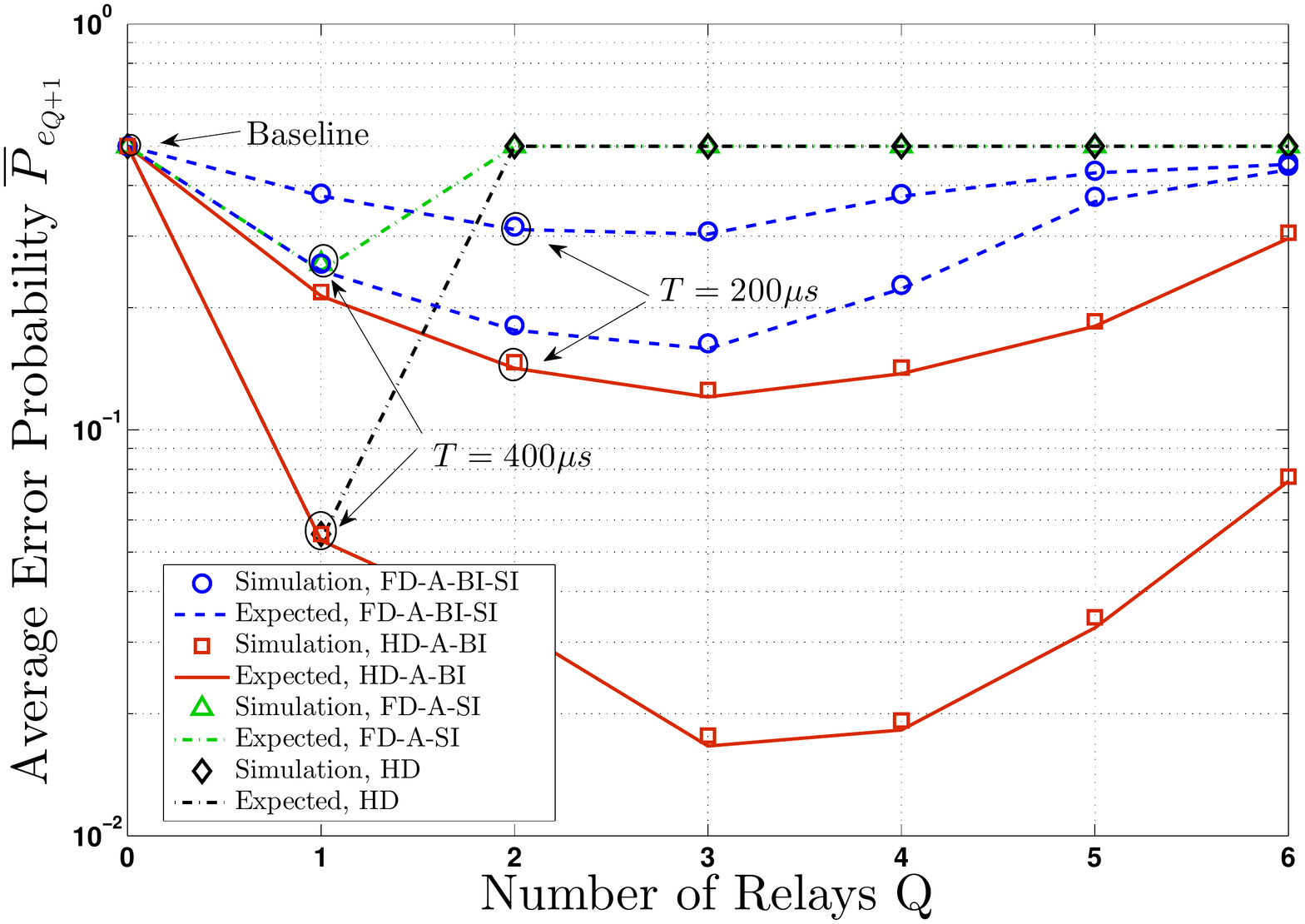}\vspace*{-6mm}
\caption{Average error probability of a SM-MH network as a function of the number of relays. $Q=0$ is the baseline case.} \label{Fig.6}
\end{figure}

In Fig. \ref{Fig.6}, we compare the performance of SM-MH with the baseline case $(Q=0)$ as a function of number of relays $Q$. We assume that $x_D = 1$ $\mu$m, $t_0=20$ $\mu$s, $M=10$, $N_{A_{1}} = 20000$, and considered the FD-A-BI-SI, HD-A-BI, FD-A-SI, and HD protocols for $T = \{ 200,400 \} $ $\mu$s. For all considered protocols, we numerically found the optimal detection threshold for the fixed part of the adaptive decision threshold that minimizes the expected error probability of the overall network. Fig \ref{Fig.6} shows that HD and FD-A-SI have the same poor performance as the baseline case when more than one relay is deployed between node $S$ and node $D$, since these two protocols cannot mitigate the backward-ISI. However, we can see that the two protocols that jointly mitigate the effects of self-interference and backward-ISI, i.e., FD-A-BI-SI and HD-A-BI, perform better than the baseline case. Furthermore, HD-A-BI performs better than FD-A-BI-SI, because the decision threshold of FD-A-BI-SI is adjusted in each bit interval based on the expected number of observed molecules to mitigate both self-interference and backward-ISI, but this expected number of molecules may differ from the actual number of observed molecules. 

A comparison of the results in Figs. \ref{Fig.2}, \ref{Fig.4}, and \ref{Fig.6} shows that for MM-MH deploying more relays is always beneficial. However, for 2M-MH and SM-MH, due to the presence of performance limiting interference, it is important to optimize the number and placement of the deployed relays.                    
\section{CONCLUSION}
\label{Sec.Con} 
In this paper, we considered a multi-hop link between nanomachines where we deployed multiple transceiver nanomachines between the transmitter and receiver nanomachines in an effort to improve the range of diffusion-based molecular communication. We considered three different relaying schemes, namely MM-MH, 2M-MH, and SM-MH. We showed both via simulation and analytical results that for 2M-MH and SM-MH the transmission of multiple random bits leads to the occurrence of self-interference, backward-ISI, and forward-ISI. We proposed two different techniques to mitigate the effect of self-interference: 1) an adaptive decision threshold at the relay, and 2) half-duplex relaying instead of full-duplex relaying. Adapting the decision threshold was also employed to mitigate the effect of backward-ISI in 2M-MH. Furthermore, we combined these methods to jointly mitigate the effects of self-interference and backward-ISI in SM-MH. In addition, we derived closed-form expressions for the expected error probability of multi-hop transmission for all considered relaying schemes. Simulation results confirmed the accuracy of the obtained error rate expressions. Our simulation and analytical results showed that the quality of communication between a transmitter nanomachine and a receiver nanomachine can be significantly improved by deploying relay nodes. 

An interesting topic for future research is the investigation of other relaying strategies such as amplify-and-forward relaying. Of interest is also the study of the impact of flow in multi-hop networks. In particular, flow may be exploited as a means for mitigation of the performance-limiting effects of self-interference and backward-ISI.   
     
\bibliographystyle{IEEEtran}
\bibliography{Library}
\end{document}